**Elastic-Viscoplastic Model for Clays: Development, Validation, and Application**

M. N. Islam[1] and C. T. Gnanendran[2]

[1]Postdoctoral Researcher, Civil and Environmental Engineering, University of Pittsburgh, 736 Benedum Hall, 3700 O'Hara St, Pittsburgh, PA 15213; Email: nislamce@gmail.com (corresponding author)

[2]Senior Lecturer, UNSW, Canberra, Australia; Email: r.gnanendran@adfa.edu.au

**Abstract**

This paper presents an elastic-viscoplastic (EVP) constitutive model in triaxial space and general stress space for isotropic clays. The EVP model is anchored in the bounding surface theory along with the mapping rule and adopts a critical state soil mechanics framework. It incorporates creep effects, and a non-linear creep function is used in the model. The EVP deformation of clay is integrated considering a reference surface and loading surface. An image parameter is deduced to establish the image surface. The strain rate tensor of the model comprises elastic-strain-rate tensor and viscoplastic-strain-rate tensor. The model formulation is capable of accounting for composite as well as single surface ellipses. Parameters of the model can be extracted from conventional oedometer and triaxial tests. The model performance is validated by capturing the behaviours in creep test, relaxation test, strain-rate effect test, and over consolidation ratio effect test of Kaolin clay, Hong Kong Marine Deposit clay, and Fukakusa clay. The model is also implemented in a Finite Element (FE) code and used to predict the long-term performance of the Nerang Broadbeach Roadway embankment constructed in Australia. The long-term settlement prediction of this embankment is also compared with that obtained with the Modified Cam Clay (MCC) model. Pertinent details of the theoretical framework of the proposed EVP model along with its validation, FE implementation and field application are discussed in this paper.

**Introduction**

Soft clays are highly compressible, exhibit both low bearing capacity and low permeability. Under the application of long-term extrinsic load, they experience immense time-dependent settlement due to their viscous behaviour (Bjerrum 1967; Graham et al. 1983). From the early 1920s to date, to illustrate the behaviour of clay, a variety of constitutive models developed, ranging from simple elastic to elastic-viscoplastic (EVP) models. Elastic-plastic constitutive models such as the Cam Clay model (Roscoe and Schofield 1963) and the Modified Cam Clay (MCC) model (Roscoe and Burland 1968) have been widely used in geotechnical engineering over the last few decades. However, such viscous-independent constitutive models are inadequate for capturing the long-term viscous behaviour of clays (Gnanendran et al. 2006; Karim et al. 2010). In addition, such behaviour of clay contributes to impeding the long-term performance of a geotechnical structure. Moreover, such behaviour often promotes progressive failure and result in very high annual maintenance costs. Hence, constitutive models accounting for the creep or time-induced deformations along with creep released pore water pressures are important to obtain a realistic explanation for the behaviour of soft clays.

In general, the time-dependent deformational behaviour of soft clays is classified under two headings: (*i*) dissipation of the excess pore-water pressure originating from the coupled hydro-mechanical process of inter-particle clay skeletons, and (*ii*) creep of soft clays which is dominated by their viscous property. On the other hand, time-dependent viscous phenomena of clays can be sub-categorized under four headings: rate dependency, creep, stress relaxation, and long-term strength (Graham et al. 1983). Herein, EVP model predictions encompassing all of those divisions are presented for different types of clays.

The formulation of the model proposed in this paper, hereafter referred as the present model, is a creep based over-stress type which is anchored in Perzyna's viscoplastic theory (Perzyna 1963). It adopts the MCC framework (Roscoe and Burland 1968), Dafalias and Herrmann mapping rule (Dafalias and Herrmann 1982, 1986), and Borja and Kavazanjian (1985) approach. However, the model proposed by Borja and Kavazanjian (1985) is formulated considering time dependent and time independent components, wherein the latter section consisted of classical plasticity theory. Here, the present model is formulated considering triaxial space as well as general stress space adopting the EVP concept. The model is validated for clays found in three places. It is also implemented in a finite element (FE) code and applied to predict long-term performance of the Nerang Broadbeach Roadway (NBR) embankment constructed in Australia.

It can be observed in the literature that the parameters required for EVP model formulation ranged between 7 (Kutter and Sathialingam 1992) to 44 (Maranini and Yamaguchi 2001). A large number of model parameters may provide a good response for a particular case, but they are not generalised. Moreover, model formulation simplicity and the determination of parameters objectively from experiments are essential for practical geotechnical application of any EVP model. In this paper, both MCC equivalent single surface as well as composite surface (Ellipses 1 and 2) based model formulation are presented; the former one requires six parameters, and latter one needs seven parameters. Five of those are identical to the MCC model parameters. The other parameters are the secondary compression index $(C_\alpha)$ and shape parameter ($R$). In composite surface based model, the parameter $R$ is essential to consider the effect of over consolidation ratio of clay. Details of the parameters are discussed in the "Model Parameters" section.

In this paper, to account for viscous effects, the concept of Borja and Kavazanjian (1985) is adopted by introducing non-linear $C_\alpha$ in the model formulation. A similar approach has also adopted in many models, such as those by Kutter and Sathialingam (1992), Hickman and Gutierrez (2007), Karim et al. (2010). However, those model formulations are different. Moreover, in the present EVP model derivation, careful consideration is essential to implement Borja-Kavazanjian concept. Otherwise, the model will be flow rule independent, which violates the theory of plasticity or might cause it to exhibit singularity problems, such as those found by Karim et al. (2010). Again, sometimes simplifications of Borja and Kavazanjian (1985) concepts contradict with model prediction such as that by Kutter and Sathialingam (1992) where elastic strain component was ignored, but significant amounts of elastic deformation were evident in the prediction, which is inconsistent with its assumption. Details of such deficiencies will be discussed in "Derivation of Viscoplastic Deformation" section under the derivation of viscoplastic deformation. In this paper, such anomalies are avoided without the need to introduce any extra model parameter.

In most EVP model formulations, $C_\alpha$ is assumed to be linear which is consistent with earlier findings, e.g. Mesri and Castro (1987), as well as those of Borja and Kavazanjian (1985), Kutter and Sathialingam (1992), Gnanendran et al. (2006), Hickman and Gutierrez (2007). However, in recent investigations, such as those by Lo et al. (2013), Karim et al. (2010), and Alonso et al. (2000), long-term laboratory tests revealed that $C_\alpha$ of clay is nonlinear. In addition, linear or constant approximation may lead to misleading prediction (Yin 1999). In EVP models presented in the recent literature, the nonlinear $C_\alpha$ assumption is either tied to a specific type of EVP model (e.g. Yin 1999) or specific type of soil, which requires fitting parameters (e.g. Karim et al. 2010). In this paper, a generalised non-linear $C_\alpha$ function is presented following the concept presented by Nash (2001) and this concept is similar to those proposed by Murakami (1979) and Yin et al. (2015). However, the function does not require any fitting parameter, nor is it tied to any specific model. In addition, comparisons of the model predictions for linear and non-linear $C_\alpha$ assumptions are also presented considering the long-term field performance data of the NBR embankment. The detail of $C_\alpha$ evaluation are also discussed in the "Model Parameter" section.

It is interesting that in most of cases, EVP models adopt a von Mises type criterion and circular yield surface in the $\pi$-plane. However, for pressure sensitive geomaterials, such a type of yield surface is not appropriate because it predicts high friction angles in a triaxial extension. In addition, failure of geomaterials is not correctly presented by a von Mises type criterion. To introduce viscous effects in EVP model formulation, adopting Borja-Kavazanjian concept along with von Mises type criterion is also bounded by similar limitation of circular shape surface in the $\pi$-plane. Such a limitation is inherent in EVP models proposed by Kutter and Sathialingam (1992) and Karim et al. (2010) also. However, both single as well as composite surface based non-circular yield surfaces are adopted in the present model, whereas in the Borja-Kavazanjian and Hickman-Gutierrez model's yield surface are MCC equivalent single surface. On the other hand, the present model is developed for clay, but the Hickman-Gutierrez model was formulated for chalk.

In addition, viscoplastic strain rate determination in the present model and the Hickman-Gutierrez model are also different, which will be discussed in next section. In the present model, to obtain realistic non-circular type surface in $\pi$-plane, the concept presented by Prashant and Penumadu (2005) is implemented, whereas in Hickman and Gutierrez (2007), it is obtained considering William and Warnke (1975) concept. In literature, there are other techniques to introduce non-circular type surface in the $\pi$-plane; among them, the most popular approaches are: (1) combining critical state theory with the Mohr-Coulomb criterion (Zienkiewicz et al. 1975) or Lade's criterion (Yao and Sun 2000), (*ii*) "transformed stress" approach considering the spatially mobilized plane (SMP) criterion (Matsuoka and Nakai 1974) or Yao et al.'s (2015) proposed Transformed Stress Method (TSM); and (*iii*) changing the definition of the Critical State line either introducing Lode Angle (Sheng et al. 2000) or *b*-value $\left(=\dfrac{\sigma_2-\sigma_3}{\sigma_1-\sigma_3}\right)$ (Prashant and Penumadu 2005). However, limitations of Zienkiewicz et al.'s (1975) approach have been discussed by Sheng et al. (2000), Matsuoka and Sun (2006). Differences among the MCC framework, TSM and SMP have been discussed by Matsuoka et al. (1999), Yao and Sun (2000), Matsuoka and Sun (2006), and Yao and Wang (2014).

Recently, Yao et al. (2015) presented an EVP model considering TSM for normally consolidated clay as well as over consolidated clay and its failure surface in $\pi$-plane is also non-circular. However, TSM is anchored in the SMP based "transformed stress" concept. In addition, a potential failure ratio is introduced in Yao et al. (2015) considering the Hvorslev line to capture over consolidated clay's peak stress ratio. In the present paper, however, the composite surface concept is used to capture the behavior of over consolidated clay without any extra model parameter. Furthermore, the present model's formulation is simple to implement in any FE code or practical geotechnical engineering application.

The non-circular surface in $\pi$-plane is achieved by changing the definition of the critical state line (CSL) slope ($M$), which is a function of *b*-value and angle of internal friction at failure. For triaxial compression and extension test, the magnitude of *b*-value is equal to 0 and 1 respectively. For compression and extension tests, $M$ is presented as $M_{b=0} = M_c$ and $M_{b=1} = M_e$ respectively. By changing the *b*-value, any stress path can be obtained. To introduce the modified form of *M* in terms of *b*-value, the method presented by Prashant and Penumadu (2005) is adopted which is also close to that of Sheng et al. (2000). The details will be presented along with the comparison with the true triaxial text experimental data and model prediction in the next section. It is worth to mentioning that Prashant and Penumadu (2005) obtained the definition of *M* from true triaxial tests result, and it was recently also adopted in Kaliakin and Leal (2013), Xiao et al. (2016), who used it successfully predict geomaterial's behavior.

The model performance is validated by capturing creep tests, relaxation tests, strain-rate effect tests, and overconsolidation ratio effect test for Kaolin clay, Hong Kong Marine Deposit (HKMD) clay, and Fukakusa clay. The model is also implemented in a code named a finite element numerical algorithm (AFENA) (Carter and Ballam 1995). The model is then applied to predict the long-term performance of surcharged preloaded embankment called the Nerang Broadbeach Roadway (NBR) embankment in Australia.

**Model Description**

It is evident that constitutive model which considered the classical theory of plasticity, such as the modified Cam Clay (MCC) model adopt single ellipse yield surface. However, limitations of the single surface cannot be overruled, as explained by Mroz (1967), Dafalias and Popov (1974) and Yu (2006). During the last few decades, the limitations of the single-surface model have opened up a wider research area. There are several approaches to overcome these limitations. However, the most popular two theories are (1) multisurface plasticity (Mroz 1967); and (2) the bounding surface (Dafalias and Popov 1974). The limitations of the multisurface theory compared with those of the bounding surface have been explained by Yu (2006). The bounding surface theory of Dafalias and Popov (1974) along with Perzyna's (1963) viscoplastic theory is adopted herein. In the proposed model, both single as well as composite surfaces are presented and controlled by shape parameter ($R$). The effect of $R$ is presented in Fig. 1.

*Bounding Ssurfaces of the Proposed Model*

For any loading history, the reference surface $\left(\bar{f}\right)$ and the loading surface $(f)$ are adopted as shown in Fig. 1. It is assumed that the potential surface $\left(\hat{f}\right)$ is identical to the reference surface, thereby invoking the associated flow. The reference surface $\left(\bar{f}\right)$ is homologous to Perzyna's (1963) static yield surface, and the viscoplastic strains are considered to exist inside, outside or on the reference surface. The loading surface $(f)$ represents the current stress state, and is analogous to the Perzyna's dynamic yield surface. The direction of the viscoplastic strain rate $[\dot{\varepsilon}]_{ij}^{vp}$ is actuated from the reference surface and $[\dot{\varepsilon}]_{ij}^{vp}$ is normal to it.

The shape of both reference surface and loading surface are considered to be same. Each surface consists of two ellipses (Ellipses 1 and 2), as shown in Fig. 2. Ellipses 1 and 2 are the modified surface of Dafalias and his co-workers, such as Dafalias and Herrmann (1982, 1986) and Kutter and Sathialingam (1992) respectively. The modification introduced is adopting a non-circular shape surface in the $\pi$-plane, whereas Dafalias and Herrmann (1982, 1986) and Kutter and Sathialingam (1992) used a von Mises type criterion, which results in a circular shaped yield surface on the $\pi$-plane. The limitations of a circular surface yield surface on the $\pi$-plane were discussed earlier. It is worth mentioning that in Ellipses 1 and 2 for $R = 2$, both ellipses reduce to the extended MCC model's single surface. Moreover, though the original MCC model adopts a von Mises type criterion, but considering the concept presented in this paper a non-circular shape surface in the $\pi$-plane also can be obtained, which is here named the extended MCC model.

The model's prediction for over consolidated clay in a composite ellipse is better compared with that for the single ellipse (Dafalias and Herrmann 1982, 1986). On the other hand, experimental data indicate that increases in the over consolidation ratio (OCR) should cause the strength locus for over-consolidated clay to approach the 'zero-tension line'. In addition, for normally consolidated clay, it intersects the CSL in the *p-q* plane (Atkinson 2007). To minimize this problem associated with a single-surface model, a composite bounding surface higher on the 'wet side' ellipse than 'dry side' ellipse for over-consolidated clay (Fig. 2) is introduced. In this paper, with the increase in *R,* Ellipse 1 (wet side) increases more than Ellipse 2. The magnitude of *R* can be deduced from conventional triaxial undrained compression tests, with its effect in the *p-q* plane presented in Fig. 1. Two ellipses of each surface meet at common tangents, as shown in Fig. 2, and allow control of each surface's. The mathematical formulations for reference and loading surfaces can be presented as follows:

$$\text{Reference Surface}, \bar{f} = \begin{cases} \bar{p}^2 - \dfrac{2}{R}\bar{p}_0\bar{p} - \dfrac{R-2}{R}\bar{p}_0^2 + (R-1)^2\left(\dfrac{\bar{q}}{M}\right)^2 & : Ellipse 1 \\ \bar{p}^2 - \dfrac{2}{R}\bar{p}_0\bar{p} + \left(\dfrac{\bar{q}}{M}\right)^2 & : Ellipse 2 \\ \bar{p}^2 - \bar{p}_0\bar{p} + \left(\dfrac{\bar{q}}{M}\right)^2, \ For\ R=2 & : Single\ ellipse \end{cases} \quad (1)$$

$$\text{Loading Surface}, f = \begin{cases} p^2 - \dfrac{2}{R}p_L p - \dfrac{R-2}{R}p_L^2 + (R-1)^2\left(\dfrac{q}{M}\right)^2 & : Ellipse 1 \\ p^2 - \dfrac{2}{R}p_L p + \left(\dfrac{q}{M}\right)^2 & : Ellipse 2 \\ p^2 - p_L p + \left(\dfrac{q}{M}\right)^2, \ For\ R=2 & : Single\ ellipse \end{cases} \quad (2)$$

where $M$ = CSL slope, $p_L$ and $\bar{p}_0$ = intersections of the loading and reference surfaces with positive $p$-axis respectively, $p$ and $q$ = mean and deviatoric pressures, respectively, in the loading surface, $\bar{p}$ and $\bar{q}$ = mean and deviatoric pressures, respectively, in the reference surface, and $R$ = shape parameter.

In Eqs. 1 and 2, to avoid the limitations of the von Mises criterion and to introduce non-circular surface in the $\pi$-plane, the critical state line slope ($M$) is presented as a function of the $b$-value $\left[=\dfrac{\sigma_2 - \sigma_3}{\sigma_1 - \sigma_3}\right]$ and the angle of internal friction at failure $(\varphi)$, which given by

$$M = \frac{6\sin\varphi\sqrt{1-b+b^2}}{3+(2b-1)\sin\varphi} \quad (3)$$

For each constant *b*-value test, the value of the peak deviatoric stress can be found from the experimental data, from which the value of $\sigma_1$, $\sigma_2$ and $\sigma_3$ can be calculated using the procedure presented by Matsuoka et al. (1999). In Fig. 3, comparisons of true triaxial test experimental data and the prediction from Eq. (3) are shown for Kaolin clay (Prashant and Penumadhu 2005), Granite (Kumruzzaman and Yin 2012), and Fukakusa clay (Ye at al. 2014), with both drained and undrained test results are presented. From Fig. 3, it is observed that Eq. (3) predicted the experimental data well.

In $\pi$-plane (Fig. 4), true triaxial experimental data presented by Kumruzzaman and Yin (2012) are compared with the modified form of the EVP model. In addition, model performance is also compared with the conventional von Mises circular type EVP model, Lade-Duncan criterion, Matsuoka Nakai criterion, Mohr Coulomb criterion, and Tresca Criterion. The modified EVP model prediction in the $\pi$-plane is very close to the experimental results, which also supports Lade-Duncan criterion. The details comparison are presented in Fig. 4.

In the general stress space, $p$, $q$ and $\bar{p}$, $\bar{q}$ can be defined as

For current stress state 
$$\begin{cases} p = \dfrac{\sigma_1 + \sigma_2 + \sigma_3}{3} \\ q = \left\{ \dfrac{1}{2} \left[ (\sigma_1 - \sigma_2)^2 + (\sigma_2 - \sigma_3)^2 (\sigma_3 - \sigma_1)^2 \right] \right\}^{\frac{1}{2}} \end{cases}$$
(4)

For reference stress state 
$$\begin{cases} \bar{p} = \dfrac{\bar{\sigma}_1 + \bar{\sigma}_2 + \bar{\sigma}_3}{3} \\ \bar{q} = \left\{ \dfrac{1}{2} \left[ (\bar{\sigma}_1 - \bar{\sigma}_2)^2 + (\bar{\sigma}_2 - \bar{\sigma}_3)^2 (\bar{\sigma}_3 - \bar{\sigma}_1)^2 \right] \right\}^{\frac{1}{2}} \end{cases}$$
(5)

where $(\sigma_1, \sigma_2, \sigma_3)$ and $(\bar{\sigma}_1, \bar{\sigma}_2, \bar{\sigma}_3)$ = principal effective stresses in loading surface and reference surface, respectively. The stress state $(\sigma_{ij})$ of loading surface $(f)$ and the reference surface $(\bar{f})$, stress state $(\bar{\sigma}_{ij})$ are interrelated through the mapping rule.

*Strain-Rate Tensor of Model*

The strain rate tensor comprises the elastic strain-rate tensor $[\dot{\varepsilon}]_{ij}^e$ and viscoplastic strain rate tensor $[\dot{\varepsilon}]_{ij}^{vp}$ such that

$$\dot{\varepsilon}_{ij} = [\dot{\varepsilon}]_{ij}^e + [\dot{\varepsilon}]_{ij}^{vp} \tag{6}$$

The elastic strain-rate tensor is simplified according to Hooke's law as

$$[\dot{\varepsilon}]_{ij}^e = [C]_{ijkl} [\dot{\sigma}]_{kl} \tag{7}$$

where $[C]_{ijkl}$ = fourth-order elastic moduli tensor; and $[\dot{\sigma}]_{kl}$ = effective stress tensor.

An isotropic linear-elastic material has two independent elements and its $[C]_{ijkl}$ matrix can be written as

$$[C]_{ijkl} = \begin{bmatrix} C_{11} & C_{12} & C_{13} & 0 & 0 & 0 \\ C_{21} & C_{22} & C_{23} & 0 & 0 & 0 \\ C_{31} & C_{32} & C_{33} & 0 & 0 & 0 \\ 0 & 0 & 0 & C_{44} & 0 & 0 \\ 0 & 0 & 0 & 0 & C_{55} & 0 \\ 0 & 0 & 0 & 0 & 0 & C_{66} \end{bmatrix}$$

$$C_{11} = C_{22} = C_{33} = \frac{1}{E}; C_{12} = C_{13} = C_{23} = \frac{-v}{E}; C_{44} = C_{55} = C_{66} = G;$$

$$E = \frac{3(1-2v)(1+e_0)p}{\kappa}; G = \frac{E}{2(1+v)}$$

The viscoplastic strain-rate tensor in Eq. (6) is generalized according to the viscoplastic theory of Perzyna (1963) as

$$[\dot{\varepsilon}]_{ij}^{vp} = \langle \phi(F) \rangle \frac{\partial f_d}{\partial \sigma'_{ij}} \tag{8}$$

$$\langle \phi(F) \rangle = \begin{cases} \phi(F) & : F > 0 \\ 0 & : F \leq 0 \end{cases}$$

$$F = \frac{f_d - f_s}{f_s}$$

where $\phi$ = rate sensitivity function; $\langle \, \rangle$ = Macaulay's bracket, $F$ = overstress function, which depends on the dynamic loading function $(f_d)$ and the static loading function $(f_s)$; $f_d$ = current stress state and $f_s$ = viscoplastic strain hardening. The normalized distance between the $f_d$ and $f_s$ is defined as $F$ in Perzyna's formulation. If $f_d$ is less than $f_s$ (i.e. $F \leq 0$), materials behave elastically (i.e. $[\dot{\varepsilon}]_{ij}^{vp} = 0$ ), but for $f_d > f_s$, the viscoplastic strain actuate as defined in Eq. 8. The functional form of $\phi$ can be obtained either experimentally or theoretically. The rate sensitivity function $(\phi)$ embodies the influence of the strain rate, whereas $\frac{\partial f_d}{\partial \sigma'_{ij}}$ represents the strain-rate direction on the reference surface.

*Derivation of Viscoplastic Deformation*

At any arbitrary reference time $(\bar{t})$, the state of soil element is at 'A' and the corresponding void ratio and preconsolidation pressure are $\bar{e}$ and $p_L$, respectively. With an increase in any time (*t*) greater than $\bar{t}$ , the soil state moves from 'A' to 'B' due to creep, and the

void ratio decreases from $\bar{e}$ to $e$ whereas the pre-consolidation pressure apparently increases from '$C$' to '$D$'. According to Fig. 5, the void ratio at $\bar{t}$ and $t$ can be written

$$\bar{e} = e_N - \lambda \ln p_L + \kappa \ln\left(\frac{p_L}{p}\right) \tag{9}$$

$$e = e_N - \lambda \ln \bar{p}_0 + \kappa \ln\left(\frac{\bar{p}_0}{p}\right) \tag{10}$$

where $\lambda$ and $\kappa$ = slopes of normal consolidation and the unloading-reloading lines, respectively, and $e_N$ = void ratio of the $\lambda-line$ when $p = 1$ kPa at $\bar{t}$. It is to note that $\bar{t}$ is not model parameter but an arbitrary reference time.

In Fig. 5, $\dot{e}_N$ = void ratio of the $\dot{\lambda}-line$ when $p = 1$ kPa at any time other than $\bar{t}$; $p_L$ and $\bar{p}_0$ are the intersections of the loading surface and reference surface with the positive $p$-axis, respectively, also presented in Fig. 1. The initial isotropic consolidation line ($\lambda-line$) at $\bar{t}$ represents the initial bounding surface. Then, because of creep, both magnitude and direction of the stress state changes which leads to gyrations of the loading surface as well as reference surface. After gyration, the surface is presented by $\dot{\lambda}-line$, which can be obtained from $\dot{e}_N$. Consolidation lines, such as the $\lambda-line$ and $\dot{\lambda}-line$ at different quasi-pre-consolidation pressures, are parallel to each other. Thereby, parallel lines maintain Bjerrum (1967) concept. By differentiating Eq. (10) with respect to time, $\dot{e}_N$ can be obtained.

Now, subtracting Eq. (9) from Eq. (10), one obtains

$$e - \bar{e} = (\lambda - \kappa) \ln\left(\frac{p_L}{\bar{p}_0}\right) \tag{11}$$

From Fig. 5, introducing the definition of $C_\alpha$ for $\bar{t}$ and $t$, then substituting $\alpha = \dfrac{C_\alpha}{ln10}$, the following expression can be obtain after rearranging

$$\frac{t}{\bar{t}} = \exp\left(\frac{\bar{e}-e}{\alpha}\right) \tag{12}$$

Eq. 12 is identical to the expression presented in the Borja and Kavazanjian [1985, their Eq. (53)]. However, the present model is formulated in the EVP framework, where as Borja and Kavazanjian model is presented considering a classical plasticity theory based time-independent component and time-dependent component. On the other hand, in Borja and Kavazanjian's (1985) model, a von Mises type criterion was used and $C_\alpha$ was constant. However, in this paper shape of yield surface in $\pi$-plane is non-circular and $C_\alpha$ is generalised non-linear function.

Differentiating Eq. (12) with respect to time yields

$$\frac{de}{dt} = -\frac{\alpha}{\bar{t}}\exp\left(\frac{e-\bar{e}}{\alpha}\right) \tag{13}$$

From the theory of viscoplasticity, the volumetric viscoplastic strain rate can be defined as

$$\dot{\varepsilon}_v^{vp} = -\frac{de}{dt}\frac{1}{1+e_0} \tag{14}$$

Combining Eqs. (13) and (14), the volumetric viscoplastic strain rate given by

$$\dot{\varepsilon}_v^{vp} = \frac{\alpha}{\bar{t}(1+e_0)}\exp\left(\frac{e-\bar{e}}{\alpha}\right) \tag{15}$$

Substituting Eq. (11) into Eq. (15) provides the following expression for volumetric viscoplastic strain rate.

$$\dot{\varepsilon}_v^{vp} = \frac{\alpha}{\bar{t}(1+e_0)}\left(\frac{p_L}{\bar{p}_0}\right)^{\frac{\lambda-\kappa}{\alpha}} \tag{16}$$

The expression for $\dot{\varepsilon}_v^{vp}$ presented in Eq. (16) is independent of the flow rule, and it is used later to determine the rate sensitivity function, $\phi(F)$ of the model. The direct application of Eq. (16) to obtain $\dot{\varepsilon}_v^{vp}$ would result in a flow-rule independent model. Furthermore, $C_\alpha$ in Eq. (16) is non-linear.

Adachi and Oka (1982) simplified Perzyna (1963) viscoplastic increment in the general stress space. If Perzyna's dynamic loading function ($f_d$) is replaced by the reference surface ($\bar{f}$), then for the associated flow rule, the viscoplastic-strain rate increment in the general stress space and triaxial space can be written as follows:

$$[\dot{\varepsilon}]_{ij}^{vp} = \phi(F)\frac{\partial \bar{f}}{\partial \bar{\sigma}_{ij}} \tag{17a}$$

$$\dot{\varepsilon}_v^{vp} = \phi(F)\frac{\partial \bar{f}}{\partial \bar{p}} \tag{17b}$$

$$\dot{\varepsilon}_q^{vp} = \phi(F)\frac{\partial \bar{f}}{\partial \bar{q}} \tag{17c}$$

In Eqs. (17-b) and (17-c), $\dot{\varepsilon}_v^{vp}$ and $\dot{\varepsilon}_q^{vp}$ = volumetric viscoplastic strain rate and deviatoric viscoplastic strain rate, respectively.

In Eqs. (17a)- (17c), the magnitude of $\phi(F)$ can be obtain combining Eq. (17-b) for $\dot{\varepsilon}_v^{vp}$ and Eq. (16) as follows

$$\phi = \frac{\alpha}{tv_0}\left(\frac{p_L}{\bar{p}_0}\right)^{\frac{\lambda-\kappa}{\alpha}}\frac{1}{\frac{\partial \bar{f}}{\partial \bar{p}}} \tag{18}$$

where, $v_0 = 1 + e_0$

If $\phi$ represented in Eq. (18) is substituted in Eq. (17b), the model will be independent of the flow rule, a mis-interpretation observed in certain existing models (e.g. Gnanendran et al. 2006; Karim et al. 2010). To avoid these discrepancies, in the present model, $\phi$ is evaluated for one-dimensional compression test conditions and does not require any additional model parameter. For one dimensional conditions, $\dfrac{\partial \bar{f}}{\partial \bar{p}}$ is replaced by $\left(\dfrac{\partial \bar{f}}{\partial \bar{p}}\right)_0$ and Eq. (18) could be rewritten as

$$\phi = \frac{\alpha}{tv_0}\left(\frac{p_L}{\bar{p}_0}\right)^{\frac{\lambda-\kappa}{\alpha}} \frac{1}{2\bar{p}_0\left[\dfrac{1}{\xi}-\dfrac{1}{R}\right]} \tag{19a}$$

$$\left.\begin{array}{l}\xi = \dfrac{\bar{p}_0}{\bar{p}} = \dfrac{-1+(R-1)\sqrt{1+R(R-2)\left(\dfrac{\eta_0}{M}\right)^2}}{R-2} \\[2em] \xi = \dfrac{\bar{p}_0}{\bar{p}} = 1+\left(\dfrac{\eta_0}{M}\right)^2, \; for\; R=2 \end{array}\right\} \tag{19b}$$

$$\left.\begin{array}{l}\eta_0 = \dfrac{\left(6(R-1)^2(\lambda-\kappa)-2\sqrt{9(\lambda-\kappa)^2(R-1)^2+(2\lambda M)^2}\right)\lambda M^2}{9(\lambda-\kappa)^2(R^4-4R^3+5R^2-2R)-(2\lambda M)^2} \\[2em] \eta_0 = -\dfrac{6(\lambda-\kappa)-2\sqrt{9(\lambda-\kappa)^2+(2\lambda M)^2}}{4\lambda}, \; For\; R=2\end{array}\right\} \tag{19c}$$

The calculation procedure for one-dimensional condition is presented by Yu (2006). To resolve the flow-rule-independent problem in EVP model formulation as discussed earlier, Hickman and Gutierrez (2007) proposed "axial scaling" concept, where the viscoplastic strain rate $[\dot{\varepsilon}]_{ij}^{vp}$ in the general stress space was calculated under triaxial compression condition using axial viscoplastic strain rate $\left(\dot{\varepsilon}_{1,TC}^{vp}\right)$.

In triaxial space, the mathematical formulation for volumetric viscoplastic strain $\left(\dot{\varepsilon}_v^{vp}\right)$ and deviatoric viscoplastic strain rate $\left(\dot{\varepsilon}_q^{vp}\right)$ can be obtain by combining Eqs. 17b, 17c, 19a and differential form of Ellipses 1 and 2 in Eq. (1) with respect to the mean pressure ($p$) and deviatoric pressure ($q$) as follows:

$$\dot{\varepsilon}_v^{vp} = \frac{\alpha_0}{\overline{tv}_0}\left(\frac{p_L}{\overline{p}_0}\right)^{\frac{\lambda-\kappa}{\alpha}} \frac{1}{\overline{p}_0\left[\frac{1}{\xi}-\frac{1}{R}\right]}\left(\overline{p}-\frac{\overline{p}_0}{R}\right) \quad ; for\ ellipse\ 1\ and\ ellipse\ 2 \tag{20-a}$$

$$\left.\begin{array}{l} \dot{\varepsilon}_q^{vp} = \dfrac{\alpha_0}{\overline{tv}_0}\left(\dfrac{p_L}{\overline{p}_0}\right)^{\frac{\lambda-\kappa}{\alpha}} \dfrac{1}{\overline{p}_0\left[\dfrac{1}{\xi}-\dfrac{1}{R}\right]}(R-1)^2\dfrac{\overline{q}}{M^2} \quad ; for\ ellipse\ 1 \\[2em] = \dfrac{\alpha_0}{\overline{tv}_0}\left(\dfrac{p_L}{\overline{p}_0}\right)^{\frac{\lambda-\kappa}{\alpha}} \dfrac{1}{\overline{p}_0\left[\dfrac{1}{\xi}-\dfrac{1}{R}\right]}\dfrac{\overline{q}}{M^2} \quad ; for\ ellipse\ 2 \end{array}\right\} \tag{20-b}$$

The viscoplastic strain rate in the general stress space for Ellipses 1 and 2 for any stress state is presented in Appendix I. To obtain the gradient matrix $[H]^n$, $[\dot{\varepsilon}]_{ij}^{vp}$ is differentiated with respect to stress state $(\sigma_{ij})$, which also explained in Appendix I. In Appendix I, to avoid repetition of similar equations, the $[H]^n$ is presented for Ellipse 1. The $[H]^n$ matrix is then included in the University of New South Wales (UNSW), Canberra modified version of finite element code named AFENA (Carter and Balaam 1995) to obtain the incremental viscoplastic stress and strain. For coupled consolidation analysis, the mathematical formulation of the load increment can be obtained by implementing the principle of virtual work to the equation of equilibrium as presented by Oka et al. (1986). The derivation of $\overline{p}_0$ and image parameter $(\beta_1)$ are presented in the Appendix II and Appendix III, respectively.

**Model Parameters**

To predict the behaviour of soil using the present model requires seven parameters for a composite ellipse: consolidation parameters (λ and κ), strength parameter (φ or M), elastic property or Poisson's ratio (ν), void ratio ($e_N$) at unit mean pressure, creep parameter $(C_\alpha)$ and shape parameter (R). In order for consolidated undrained triaxial tests results to predict the pore water pressure additional parameter, permeability of the clay $(K_i)$ is essential. For a single ellipse, R = 2 and the model requires six parameters.

The consolidation parameters are the gradient of the normal consolidation line $(\lambda)$ and the gradient of swelling line $(\kappa)$ which can be obtained from isotropic compression tests or the compression index $(C_c)$ and the swelling index $(C_s)$ from conventional one-dimensional consolidation tests $\left(\lambda = \dfrac{C_c}{2.303}, \kappa = \dfrac{C_s}{2.303}\right)$. The strength parameter is determined either from the slope of the critical state line or angle of internal friction $(\varphi)$. The void ratio $(e_N)$ at the unit mean pressure at any arbitrary reference time $(\bar{t})$ can be obtained from one-dimensional consolidation tests, and the change in the void ratio $(\dot{e}_N)$ can be found from the differentiation of Eq. (10).

To account viscosity of clay, non-linear $C_\alpha$ is introduced in the model, which is similar to Nash (2001) and given by

$$\frac{C_{\alpha i}}{C_{\alpha i-1}} = \left(\frac{p_i}{p_{i-1}}\right)^{\frac{\lambda-\kappa}{\alpha_{i-1}}} \tag{21}$$

where, $p_i$ in this model is referred with respect to $\bar{p}_0$. The value of $C_\alpha$ can be obtained from the oedometer test or triaxial test.

The shape parameter $(R)$ controls the yield function shape. With the increase of $R$, the shape of the surface becomes flat. There are several proposals for determining the magnitude of $R$, such as that of Dafalias and Herrmann (1986) in which, for normally consolidated clay, an analytical expression of the undrained stress path is presented as

$$\frac{|\bar{q}|}{\bar{p}_0} = \frac{M}{R-1}\left[\frac{2}{R}\left(\frac{\bar{p}}{\bar{p}_0}\right)^{\frac{\lambda-2\kappa}{\lambda-\kappa}} + \left(1-\frac{2}{R}\right)\left(\frac{\bar{p}}{\bar{p}_0}\right)^{\frac{-2\kappa}{\lambda-\kappa}} - \left(\frac{\bar{p}}{\bar{p}_0}\right)^2\right]^{\frac{1}{2}} \tag{22}$$

For any given values of $\lambda$, $\kappa$, $M$ and $\bar{p}_0$, the magnitude of $R$ can be determined by fitting the data obtained from the undrained stress path. An alternate empirical approach to predict the magnitude of $R$ is also available in Islam (2014) and both provide identical value of $R$.

**Validation of the Model and Discussion**

The model presented in this paper is applied to predict consolidated undrained triaxial compression and extension tests, consolidated drained compression tests, overconsolidation ratio effect tests, confining pressure effects, strain-rate effect tests, creep tests, and relaxation tests. This verification includes Kaolin clay (Herrmann et al. 1982), HKMD clay (Yin and Zhu 1999; Yin et al. 2002) and Fukakusa clay (Adachi and Oka 1982). The clay properties are presented in Table 1.

*Simulations of Consolidated Undrained Triaxial Tests of Kaolin Clay*

Herrmann et al. (1982) conducted extensive undrained triaxial compression and extension tests on reconstituted Kaolin clay [liquid limit (LL) = 47% and plasticity index (PI) = 27%] for different over-consolidation ratios (OCRs). Comparisons of the measured and predicted deviatoric stress versus axial strain responses from the model considering different OCRs for consolidated undrained triaxial compression (for OCR = 1, 2, 4 and 6) and undrained triaxial extension (for OCR = 1 and 2) are shown in Fig. 6(a). It is evident that, for normally consolidated soil (OCR=1), before the peak deviatoric stress, the model captured the stress–strain response well, but after the peak stress, the model slightly underpredicted it. The difference between the predicted and experimental data is approximately 3.5% near the peak deviatoric stress, which decreased with increases in the strain; at 14% strain, the under-prediction is only 1.4%. In Fig. 6(a), for OCR = 2, 4 and 6, before attaining the peak stress, the model over-predicted. But, afterwards it exhibited only small magnitudes of under-prediction. Similar trends are also observed in the results from the consolidated undrained triaxial extension tests.

In Fig. 6(b), a comparison of the experimental and predicted stress paths are presented in which it is evident that, for normally consolidated soil (OCR =1), the model's predictions are satisfactory for both consolidated undrained triaxial compression and consolidated undrained triaxial extension tests. However, for OCR = 2, the model slightly under-predicted the stress path. For over-consolidated soils, such as OCR = 4 and 6, the predictions for compression tests are similar and in the overall sense satisfactory.

In Fig. 6©, comparison of the experimental and predicted responses of the pore-water pressures for consolidated undrained triaxial compression and consolidated undrained triaxial extension tests are presented considering OCR = 1 and 2. For normally consolidated soil, the predictions for both compression and extension tests are satisfactory, although a small amount of under-prediction is noticeable. For OCR = 2, in the triaxial compression test, the model slightly overpredicted before the maximum pore-water pressure is reached. In the triaxial extension test prediction for OCR=2, the negative pore-water pressure is well captured with a small magnitude of underprediction.

*Simulations of Consolidated Drained and Undrained Triaxial Tests on HKMD Clays*

The performance of model predictions conducted on HKMD clays (Yin and Zhu (1999), and Yin et al. (2002)) was also assessed. This is a reconstituted medium plastic clay (LL=60% and PI= 32%). In this paper, isotropic consolidated drained triaxial tests of normally consolidated clay for different mean pressures and isotropic normally consolidated undrained stage-changed strain rate with relaxation, and creep tests are presented.

*Simulations of Drained Tests on HKMD Clay*

In this section, the predicted isotropic consolidated drained test behaviour of normally consolidated HKMD clay considering mean pressures of 300 kPa and 400 kPa is discussed. Although drained shear tests are not frequently performed on clay in a laboratory, but such test conditions are evident in certain field cases. Therefore, investigation of drained shear test is important. In Figs. 7(a-c), comparisons of the measured (Yin and Zhu 1999) and predicted drained responses of normally consolidated HKMD clay are presented, which shows that the stiff behavior exhibited by the normally consolidated clay was captured well by the model. In Fig. 7(a), it can be observed that the model captured the deviatoric stress versus axial strain responses up to a 7.5% axial strain well, and then marginally under-predicted them by approximately 3.3 %. The volumetric responses and axial strains shown in Fig. 7(b) and stress path predictions in Fig. 7(c) indicate that the model's predictions are very close to the experimental measurements.

*Simulations of Undrained triaxial tests at stage changed strain rate on HKMD Clay*

Yin et al. (2002) conducted a stage-changed axial strain rate isotropically consolidated undrained triaxial compression tests on normally consolidated HKMD clay. The mean pressure and back pressure are, respectively, 300 kPa and 200 kPa. The axial strain rate at different stages is different. In some cases, without applying a strain rate and maintaining constant mean pressure, the deviatoric stress decreased with time. Even though the loading history is complicated, the model captured experimental results with success. The deviatoric stress versus axial strain, pore pressure and stress path responses are presented in Figs. 8(a-c), respectively.

*Simulations of Undrained triaxial creep tests on HKMD Clay*

Zhu (2000) conducted three consolidated-undrained creep tests on HKMD clays. Initially three samples were normally and isotropically consolidated with 400 kPa pressure, then deviatoric stress of 134, 189, and 243 kPa was applied instantly. In Figs 9 (a and b), axial strain-time relation and pore pressure-time relation are predicted. It is observed from Fig. 9 that at a low stress level, the model captured well the experimental results which are also identical to the observed response presented by Yin et al. (2002) and Yao et al. (2015).

*Simulations of Undrained triaxial tests at various strain rates on Fukakusa Clay*

Adachi and Oka (1982) conducted extensive undrained triaxial compression tests on reconstituted Fukakusa clay (LL = 48.5%, and PI = 21.8%) for strain rates of 0.0835%/min and 0.00817%/min. From Figs. 10(a and b), it is evident that model captured well the experimental data. Some discrepancies were observed close to the critical state line, and similar predictions for strain rate test on HKMD clay are also available in Yao et al. (2015).

**Application of the Model**

The model in this paper was adopted to predict the long-term performances of the Nerang Broad-beach Roadway (NBR) embankment located close to the Gold Coast Highway in the southern part of Surfers Paradise, Gold Coast, Queensland, Australia. The NBR embankment was founded on deep Cainozoic estuarine alluvial soft sensitive clay deposits of thicknesses from 5.0 to 21.0 m overlaying greywacke and argillite bedrock. The length of the embankment is approximately 1.3 km and its width varies between 20.00 m and 28.00 m. To reduce the post-construction settlement, preloading was conducted along the roadway in 28 different sections. To monitor the long-term performance of the embankment, a total 18 settlement plates and four piezometers were placed. To delineate the subsurface profile of the embankment, extensive investigations were carried out (Main Roads 2001) including six borehole, 20 cone penetration tests (CPT) and four piezocone dissipation tests (CPT-u) tests. Undisturbed samples were collected from different depths for comprehensive laboratory experiments, including Atterberg limits, triaxial, moisture content, density, particle size distribution and oedometer consolidation tests. The details of the geology, subsurface profile, geotechnical properties, construction history, longitudinal section and cross-section, instrumented monitored data have been given by Islam et al. (2013, 2015) and Islam (2014).

In this paper, one embankment section observed behaviour was predicted using the model and compared with the MCC model's prediction. The filling height (H) of the embankment consisted of 3.0 m preloading and 1.0 m surcharging. The preloading was monitored for 370 days while surcharging was additionally monitored for 220 days. To avoid the stability problem, the surcharging was applied rather than full height of the embankment. Table 2 presents the model parameters for different ranges of reduced level (RL), which were obtained from the interpretation of the laboratory obtained data along with the cone penetration tests (CPT) and piezocone dissipation tests (CPT-u) data. Details of the parameters determination process have been given by Islam (2014) and Islam et al. (2015).

The embankment section for the model prediction is shown in Fig. 11, which consists of 24,813 nodes and 12,240 six noded triangle elements. From Fig. 12, it is evident that for preloading and surcharging the MCC model captured measured settlements until construction time (60 days) of embankment and then started to deviate from the measured settlement. For 3.0 m preloading after 370 days, the MCC model under predicted by 13.30 % and after 590 days the magnitude of under prediction for surcharging was 14.25%. This might be attributable to the ongoing creep of the soil, which was ignored in the MCC model. For the EVP model, it was observed that for linear and non-linear $C_\alpha$ cases, the model had better predictions compared with the MCC model. However, the non-linear $C_\alpha$ prediction is much closer to the field observation than those of linear $C_\alpha$.

For the preloading section, after 370 days the EVP model under predictions for linear and non-linear $C_\alpha$ are 5.9 and 3.45 %, respectively, whereas underprediction during surcharging after 590 days are 5.3 and 3.2%, respectively. For the Leneghans embankment in Australia, Karim et al. (2010) also reported that non-linear $C_\alpha$ based EVP model better predicted the field response. However, Karim et al. (2010) proposed non-linear $C_\alpha$ requires fitting parameters. Lo et al. (2013) reported that Yin (1999) proposed non-linear $C_\alpha$ is limited to the specific model presented by Yin and his Co-workers like Yin (1999). In this paper, a generalised non-linear $C_\alpha$ is presented.

**Concluding remarks**

In this paper, a two-surface EVP model is formulated in general stress space as well as in triaxial space to describe the time-dependent viscous behavior of clays. The model was validated and implemented using a FE code. A non-linear creep function was also adopted. The yield surface adopted in the model is analogous to the MCC model and isotropic state is invoked. Hence, the present model might not be suitable for cases where the clay is anisotropically consolidated or consideration of rotation of principal stress direction is essential. To resolve these limitations of the model in the current framework, extra model parameters will be required. The primary concerns of the proposed model are the number of model parameters, simplicity of the derivation, formulation in general stress space and prediction performance. The framework of the proposed model is capable to consider composite ellipse surface along with single surface, while former requiring seven and latter six parameters. Because of the simplicity of the model formulation and parameters extraction, this model might be useful for practical engineering application.

The proposed model was used to simulate the experimental results of triaxial tests for Kaolin clay, HKMD clay, and Fukakusa clay reported in the literature. The model captured a wide range of experiments under drained and undrained conditions considering triaxial compression and extension tests, creep tests, relaxation tests, and strain rate tests. From the model predictions, it was observed that the model slightly over predicted the non-linear response at small strain level. This might be overcome by introducing hysteretic response equation considering Whittle and Kavvadas (1994), but additional model parameters would be required.

The proposed EVP model is applied to predict the long-term performance (590 days) of the NBR embankment. It is evident from the comparison of the predicted performance with the field data that the EVP model captured the time-dependent behaviour well than elastic-plastic model such as the MCC model. On the other hand, implementation of linear or non-linear $C_\alpha$ in EVP model impacted the prediction performances, and latter captured the field response more closely than the former.

**Appendix I: Derivation of Gradient Matrix**

The viscoplastic strain rate for associated flow rule for composite surface can be written as follows

$$\dot{\varepsilon}_{ij}^{vp} = \phi(F) \frac{\partial \bar{f}}{\partial \bar{\sigma}_{ij}} \tag{23}$$

$$\frac{\partial \bar{f}}{\partial \bar{\sigma}_{ij}} = \frac{\partial \bar{f}}{\partial \bar{p}} \frac{\partial \bar{p}}{\partial \bar{\sigma}_{ij}} + \frac{\partial \bar{f}}{\partial \bar{q}} \frac{\partial \bar{q}}{\partial \bar{\sigma}_{ij}} + \frac{\partial \bar{f}}{\partial M} \frac{\partial M}{\partial b} \frac{\partial b}{\partial \bar{\sigma}_{ij}}$$

$$\frac{\partial \overline{f}}{\partial \overline{p}} = 2\left[\overline{p} - \frac{\overline{p}_0}{R}\right] \quad ; \text{for ellipse 1 and ellipse 2}$$

$$\frac{\partial \overline{f}}{\partial \overline{q}} = 2(R-1)^2 \frac{\overline{q}}{M^2} \quad ; \text{for ellipse 1}$$

$$= \frac{2\overline{q}}{M^2} \quad ; \text{for ellipse 2}$$

$$\frac{\partial \overline{p}}{\partial \overline{\sigma}_{ij}} = \frac{1}{3}\delta_{ij} \qquad \delta_{ij} = \begin{cases} 1 & \text{if } i = j \\ 0 & \text{if } i \neq j \end{cases}$$

$$\frac{\partial \overline{q}}{\partial \overline{\sigma}_{ij}} = \begin{cases} \dfrac{3}{2\overline{q}}(\overline{\sigma}_{ij} - \overline{p}\delta_{ij}) & \text{if } i = j \\[2mm] \dfrac{3}{2\overline{q}}(2\overline{\sigma}_{ij}) & \text{if } i \neq j \end{cases}$$

$$\frac{\partial \overline{f}}{\partial M} = \frac{-2(R-1)^2 \overline{q}^2}{M^3}; \text{for ellipse 1}$$

$$\frac{\partial \overline{f}}{\partial M} = \frac{-2\overline{q}^2}{M^3}; \text{for ellipse 2}$$

$$\frac{\partial M}{\partial b} = \frac{3\sin\varphi(2b-1)}{\left(\sqrt{b^2-b+1}\right)\left[3+(2b-1)\sin\varphi\right]} - \frac{12\sin^2\varphi\left(\sqrt{b^2-b+1}\right)}{\left[3+(2b-1)\sin\varphi\right]^2}$$

$$\frac{\partial b}{\partial \overline{\sigma}_{11}} = -\frac{\overline{\sigma}_{22} - \overline{\sigma}_{33}}{(\overline{\sigma}_{11} - \overline{\sigma}_{33})^2}$$

$$\frac{\partial b}{\partial \overline{\sigma}_{22}} = \frac{1}{\overline{\sigma}_{11} - \overline{\sigma}_{33}}$$

$$\frac{\partial b}{\partial \overline{\sigma}_{33}} = -\frac{1}{\overline{\sigma}_{11} - \overline{\sigma}_{33}} + \frac{\overline{\sigma}_{22} - \overline{\sigma}_{33}}{(\overline{\sigma}_{11} - \overline{\sigma}_{33})^2}$$

McDowell and Hau (2003) stated that for plain strain or axisymmetric condition, the third part of chain equation is negligible and to derive $H^n$ matrix for plain strain condition it is neglected. It is to note that for embankment analysis plain strain condition is assumed.

For ellipse 1:

$$\dot{\varepsilon}_{ij}^{vp} = c_1 c_2 \left(\frac{p_L}{\bar{p}_0}\right)^{\frac{\lambda-\kappa}{\alpha}} \beta_1 \begin{cases} \dfrac{2p}{3} - \dfrac{2p_L}{3R} - \dfrac{3(R-1)^2 p}{M^2} + \dfrac{3(R-1)^2 \sigma_{ij}}{M^2} & : i = j \\ \dfrac{6(R-1)^2 \sigma_{ij}}{M^2} & : i \neq j \end{cases}$$

For ellipse 2:

$$\dot{\varepsilon}_{ij}^{vp} = c_1 c_2 \left(\frac{p_L}{\bar{p}_0}\right)^{\frac{\lambda-\kappa}{\alpha}} \beta_1 \begin{cases} \dfrac{2p}{3} - \dfrac{2p_L}{3R} - \dfrac{3p}{M^2} + \dfrac{3\sigma_{ij}}{M^2} & : i = j \\ \dfrac{6\sigma_{ij}}{M^2} & : i \neq j \end{cases}$$

$$c_1 = \frac{\alpha_0}{t v_0}, \quad c_2 = \frac{1}{2\bar{p}_0 \left[\dfrac{1}{\xi} - \dfrac{1}{R}\right]}$$

The element of gradient matrix $H^n = \left(\dfrac{\partial \dot{\varepsilon}_{ij}^{vp}}{\partial \sigma_{ij}}\right)^n$ considering ellipse 1 the plane strain condition is presented here.

$$\text{Assumed}: a_1 = \left(p + \frac{q^2}{pM^2}\right); a_2 = 2p^2 + \frac{2(R-1)^2 q^2}{M^2}; a_3 = \left(\frac{p + \dfrac{q^2}{pM^2}}{\bar{p}_0}\right)^{\frac{\lambda-\kappa}{\alpha}}; a_4 = 1 - \frac{q^2}{p^2 M^2}; a_5 = \frac{2p\bar{p}_0}{R};$$

$$a_6 = \sqrt{1 + \left(1 + \frac{(R-1)^2 q^2}{p^2 M^2}\right) R(R-2)}; a_7 = \frac{2}{3} - \frac{3(R-1)^2}{M^2}; a_8 = \frac{3(R-1)^2}{pM^2}; a_9 = \frac{2}{3} \frac{p + \dfrac{q^2}{pM^2}}{R};$$

$$a_{10} = \frac{2}{3} \frac{1 - \dfrac{q^2}{p^2 M^2}}{R}; a_{11} = 2p^2 + \frac{2(R-1)^2 q^2}{M^2}, a_{12} = \frac{2\bar{p}_0}{R}; a_{13} = \frac{2\bar{p}_0 (R-1)^2 (R-2) q^2}{p^2 M^2}$$

$$For, i = j, \frac{\partial \dot{\varepsilon}_{ij}^{vp}}{\partial p} = \frac{1}{\alpha a_1 a_2}\left(c_1 c_2 a_3 (\lambda - \kappa) a_4 (a_5 + a_5 a_6)\left(p(a_7 + a_8 \sigma_{ij}) - a_9\right)\right)$$

$$+ \frac{1}{a_{11}}\left(c_1 c_2 a_3 \left(a_{12} + a_{12} a_6 - \frac{a_{13}}{a_6}\right)\left(p(a_7 + a_8 \sigma_{ij}) - a_9\right)\right) \quad (24)$$

$$+ \frac{1}{a_{11}}(c_1 c_2 a_3)(a_5 + a_5 a_6)(a_7 - a_{10}) - \frac{1}{a_{11}^2}\left(4 c_1 c_2 a_3 (a_5 + a_5 a_6)\left(p(a_7 + a_8 \sigma_{ij}) - a_9\right)p\right)$$

$$For, i = j, \frac{\partial \dot{\varepsilon}_{ij}^{vp}}{\partial q} = \frac{1}{pM^2 \alpha a_1 a_2}\left(2 c_1 c_2 a_3 (\lambda - \kappa) q (a_5 + a_5 a_6)\left(p(a_7 + a_8 \sigma_{ij}) - a_9\right)\right)$$

$$+ \frac{1}{pM^2 a_6 a_{11}}\left(2 c_1 c_2 a_3 \bar{p}_0 q (R-1)^2 (R-2)\left(p(a_7 + a_8 \sigma_{ij}) - a_9\right)\right) \quad (25)$$

$$- \frac{4}{3}\frac{c_1 c_2 a_3 (a_5 + a_5 a_6) q}{pM^2 R a_{11}} - \frac{1}{a_{11}^2 M^2}\left(4 c_1 c_2 a_3 (a_5 + a_5 a_6)\left(p(a_7 + a_8 \sigma_{ij}) - a_9\right)(R-1)^2 q\right)$$

$$For, i \neq j, \frac{\partial \dot{\varepsilon}_{ij}^{vp}}{\partial p} = \frac{1}{M^2 \alpha a_1 a_2}\left(6 c_1 c_2 a_3 (\lambda - \kappa) a_4 (a_5 + a_5 a_6)(R-1)^2 \sigma_{ij}\right)$$

$$+ \frac{1}{M^2 a_2}\left(6 c_1 c_2 a_3 \left(a_5 + a_5 a_6 - \frac{2\bar{p}_0 (R-1)^2 (R-2) q^2}{p^2 M^2 a_6}\right)(R-1)^2 \sigma_{ij}\right) \quad (26)$$

$$- \frac{1}{M^2 a_2^2}\left(24 c_1 c_2 a_3 (a_5 + a_5 a_6) p (R-1)^2 \sigma_{ij}\right)$$

$$For, i \neq j, \frac{\partial \dot{\varepsilon}_{ij}^{vp}}{\partial q} = \frac{1}{\alpha p M^4 a_1 a_2}\left(12 c_1 c_2 a_3 (\lambda - \kappa) q (a_5 + a_5 a_6)(R-1)^2 \sigma_{ij}\right)$$

$$+ \frac{1}{pM^4 a_2 a_6}\left(12 c_1 c_2 a_3 \bar{p}_0 q (R-1)^4 (R-2) \sigma_{ij}\right) \quad (27)$$

$$- \frac{1}{a_2^2 M^4}\left(24 c_1 c_2 a_3 (a_5 + a_5 a_6) q (R-1)^4 \sigma_{ij}\right)$$

## Appendix II: Derivation of Hardening Rule

Differentiating equation 11 with respect to time and re-arranging it for the incremental creep-inclusive pre-consolidation pressure $(d\bar{p}_0)$, it can be written as

$$\frac{d\bar{p}_0}{dt} = p_L \exp\left(\frac{\bar{e} - e}{\lambda - \kappa}\right)\left(\frac{-de}{dt}\right)\frac{1}{\lambda - \kappa} \quad (28)$$

Now, substituting equation 13 in equation 28, we obtain

$$\frac{d\overline{p}_0}{dt} = p_L \frac{\alpha}{\overline{t}(\lambda-\kappa)} \exp\left(\frac{\overline{e}-e}{\lambda-\kappa}\right) \exp\left(\frac{e-\overline{e}}{\alpha}\right) \tag{29}$$

Then, substituting equation 11 in equation 29, this expression becomes

$$\frac{d\overline{p}_0}{dt} = \overline{p}_0 \frac{\alpha}{\overline{t}(\lambda-\kappa)} \left(\frac{p_L}{\overline{p}_0}\right)^{\frac{\lambda-\kappa}{\alpha}} \tag{30}$$

Combining equation 30 with equation 16, we obtain

$$\frac{d\overline{p}_0}{dt} = \frac{1+e_0}{\lambda-\kappa} \overline{p}_0 \dot{\varepsilon}_v^{vp} \tag{31}$$

The creep-inclusive pre-consolidation pressure $(\overline{p}_0)$ following the evolving rule can be written as

$$(\overline{p}_0)_i = (\overline{p}_0)_{i-1} + (d\overline{p}_0)_i \tag{32}$$

## Appendix III: Derivation of Image Parameter

To map the image stress of the loading surface $(p,q)$ on the reference surface $(\overline{p},\overline{q})$, the loading surface mapping parameter $(\beta_1)$ is used as

$$\left.\begin{array}{l} \overline{\sigma}_{ij} = \beta_1 \sigma_{ij} \\ \overline{p} = \beta_1 p \\ \overline{q} = \beta_1 q \end{array}\right\} \tag{33}$$

Now, substituting equation 33 in equation 1 (ellipse 1) and solving it for the quadratic real value, we obtain $\beta_1$ as

$$\left.\begin{array}{l}\beta_1 = \dfrac{\dfrac{2p\bar{p}_0}{R} + \dfrac{2p\bar{p}_0}{R}\sqrt{1 + \left[1 + \dfrac{(R-1)^2 q^2}{p^2 M^2}\right](R-2)R}}{2\left[p^2 + \dfrac{(R-1)^2 q^2}{M^2}\right]} \\ \\ \beta_1 = \dfrac{\bar{p}_0}{p + \dfrac{q^2}{pM^2}} = \dfrac{\bar{p}_0}{p_L}, For\ R = 2 \end{array}\right\}$$  (34)

**Acknowledgements**

The first author was supported by the Tuition Fee Scholarship (TFS) while conducting his Doctoral research at UNSW, Canberra, Australia. Also, the authors wish to express their gratitude to Dr Rajibul Karim, former PhD student, UNSW, Canberra, for his valuable comments and discussions.

**List of symbols**

| | | |
|---|---|---|
| $C_\alpha$ | : | Coefficient of secondary consolidation |
| $C_{ijkl}$ | : | Fourth-order elastic moduli tensor |
| $C_c$ | : | The compression index |
| $C_s$ | : | The swelling index |
| $E$ | : | Young's modulus |
| $e$ | : | Current void ratio |
| $\bar{e}$ | : | Void ratio at reference time |
| $e_N$ | : | Void ratio at reference time when $p = 1$kPa on $\lambda$-line |
| $\dot{e}_N$ | : | Void ratio at any time when $p = 1$kPa on the $\dot{\lambda}$-line |
| $F$ | : | The overstress function |
| $f$ | : | The loading surface |
| $\bar{f}$ | : | The reference Surface |
| $\hat{f}$ | : | The potential surface |
| $f_d$ | : | The dynamic loading function |
| $f_s$ | : | The static loading function |
| $G$ | : | Shear modulus |
| $K_i$ | : | Permeability of the clay |
| $M$ | : | Slope of the critical state line in $(q - p)$ space |
| $P$ | : | Mean effective stress on the $f$ surface |
| $\bar{p}$ | : | Mean effective stress on the $\bar{f}$ surface |
| $p_L$ | : | Intersection of $f$ surface with the positive $p$ axis |
| $\bar{p}_0$ | : | Intersection of $\bar{f}$ surface with the positive $p$ axis |
| $p_c$ | : | The pre-consolidation pressure |
| $q$ | : | Deviator stress on the $f$ surface |
| $\bar{q}$ | : | Deviator stress on the $\bar{f}$ surface |
| $R$ | : | Shape parameter |
| $\bar{t}$ | : | Reference time |
| $\alpha$ | : | Coefficient of secondary consolidation in the natural log scale |
| $\beta_1$ | : | The mapping parameters |
| $\sigma_{ij}$ | : | The stress state on loading surface |
| $\bar{\sigma}_{ij}$ | : | The stress state on reference surface |
| $\dot{\varepsilon}_{ij}$ | : | Total strain rate tensor |
| $\dot{\varepsilon}_{ij}^{vp}$ | : | Viscoplastic strain rate tensor |
| $\dot{\varepsilon}_{ij}^{e}$ | : | Elastic strain rate tensor |
| $\dot{\varepsilon}_{v}^{vp}$ | : | Volumetric viscoplastic strain rate |
| $\dot{\varepsilon}_{q}^{vp}$ | : | Deviatoric viscoplastic strain rate |
| $\lambda$ | : | The gradient of the normal consolidation line |
| $\kappa$ | : | The gradient of swelling line |
| $\varphi$ | : | The angle of internal friction |
| $\phi$ | : | The rate sensitivity function |
| $\nu$ | : | The poisson's ratio |
| $\nu_0$ | : | The specific volume |
| $\bar{\eta}_0$ | : | The stress ratio, $\dfrac{\bar{q}}{\bar{p}} = \dfrac{q}{p}$ |
| $\xi$ | : | The equivalent stress ratio, $\dfrac{\bar{p}_0}{\bar{p}}$ |

**References**

Adachi, T., and Oka, F. (1982). "Constitutive equations for normally consolidated clay based on Elasto-viscoplasticity." *Soils and Foundations*, 22(4), 57-70.

Alonso, E. E., Gens, A., and Lloret, A. (2000). "Precompression design for secondary settlement reduction." *Geotechnique*, 50(6), 645-656.

Atkinson, J. (2007). "Peak strength of overconsolidated clays." *Geotechnique*, 57(2),127-135.

Bjerrum, L. (1967). "Engineering geology of Norwegian normally consolidated marine clays as related to settlements of buildings." *Geotechnique*, 17(2), 81-118.

Borja, R. I., and Kavazanjian, J. E. (1985). "A constitutive model for the stress-strain&-time behaviour of 'wet' clays." *Geotechnique*, 35(3), 283-298.

Carter, J. P., and Balaam, N. P. (1995). AFENA User's Manual. Version 5.0, Center for Geotechnical Research, University of Sydney, Sydney-2006, Australia.

Dafalias, Y. F., and Popov, E. P.(1974). "A model of non-linearly hardening materials for complex loadings." Proc., Proceedings of the 7th US National Congress of Applied Mechanics, 173-192.

Dafalias, Y. F., and Herrmann, L. R. (1982). "Bounding surface formulation of soil plasticity Soil Mechanics- Transient and Cyclic Loads." *Soil Mechanics- Transient and Cyclic Loads,* Editors Pande, G. N. and Zienkiewicz, O. C.Wiley, Chichester: 253-282.

Dafalias, Y. F., and Herrmann, L. R. (1986). "Bounding Surface Plasticity. II: Application to Isotropic Cohesive Soils." *Journal of Engineering Mechanics*, 112(12), 1263-1291.

Gnanendran, C. T., Manivannan, G., and Lo, S. C. R. (2006). "Influence of using a creep, rate, or an elastoplastic model for predicting the behaviour of embankments on soft soils." *Canadian Geotechnical Journal*, 43(2), 134-154.

Graham, J., Crooks, J. H. A., and Bell, A. L. (1983). "Time effects on the stress-strain behaviour of natural soft clays." *Geotechnique*, 33(3), 327-340.

Herrmann, L. R., Shen, C. K., Jafroudi, S., DeNatale, J. S., and Dafalias, Y. F. (1982). "A verification study for the bounding surface plasticity model for cohesive soils." *N. C. B. C. Final report to the Civil Engineering Laboratory*, Port Hueneme, California.

Hickman, R. J., and Gutierrez, M. S. (2007). "Formulation of a three-dimensional rate-dependent constitutive model for chalk and porous rocks." *International Journal for Numerical and Analytical Methods in Geomechanics*, 31(4), 583-605.

Islam, M. N., Gnanendran, C. T., Sivakumar, S. T., and Karim, M. R. (2013). "Long-term performance of a preloaded road embankment." 1*8th ICSMGE*, D. Pierre, D. Jacques, F. Roger, P. Alain, and S. Francois, eds. Paris, France, 1291-1294.

Islam, M. N. (2014). "Associated and non-associated flow rule based elastic-viscoplastic models for clays." *PhD Thesis*, UNSW, Canberra, Australia.

Islam, M. N., Gnanendran, C. T., and Sivakumar, S. T. (2015). "Prediction of embankment time-dependent behaviour of on soft soils: effects of preloading, surcharging and the choice of lab versus field test data for soft soil parameters." *Ground Improvement Case Histories*, volume 1, Chapter 13, B. Indraratna, Chu, J. and Cholachat, ed., 359-379.

Kaliakin, V., and Leal, A. N. (2013). "Investigation of critical states and failure in true triaxial tests of clays." *Constitutive Modeling of Geomaterials*, Springer, 185-191.

Karim, M. R., Gnanendran, C. T., Lo, S. C. R., and Mak, J. (2010). "Predicting the long-term performance of a wide embankment on soft soil using an elastic–viscoplastic model." *Canadian Geotechnical Journal*, 47(2), 244-257.

Kumruzzaman, M., and Yin, J.-H. (2012). "Influence of the intermediate principal stress on the stress–strain–strength behaviour of a completely decomposed granite soil." *Géotechnique*, 62(3), 275-280.

Kutter, B. L., and Sathialingam, N. (1992). "Elastic-viscoplastic modelling of the rate-dependent behaviour of clays." *Geotechnique*, 42(3), 427 –441.

Lo, S. R., Karim, M. R., and Gnanendran, C. T. (2013). "Consolidation and Creep Settlement of Embankment on Soft Clay: Prediction Versus Observation." *Geotechnical Predictions and Practice in Dealing with Geohazards*, J. Chu, S. P. R. Wardani, and A. Iizuka, eds., Springer Netherlands, 77-94.

Main Roads. (2001). "Additional Geotechnical Investigation for the Proposed Western RSS Wall Area, Nerang-Broadbeach Deviation, Gooding's Corner." *Queensland Department of Transport, Report: R3233*, Australia.

Maranini, E., and Yamaguchi, T. (2001). "A non-associated viscoplastic model for the behaviour of granite in triaxial compression." *Mechanics of Materials*, 33(5), 283-293.

Matsuoka, H., and Nakai, T. (1974). "Stress deformation and strength characteristics of soil under three different principal stresses."*Proceeding of Japanese Society of Civil Engineering*, 59-70.

Matsuoka, H., Yao, Y.-P., and Sun, D. A. (1999). "The Cam Clay Models Revised by the SMP Criterion." *Soils and Foundations*, 39(1), 81-95.

Matsuoka, H., and Sun, D. A. (2006). *The SMP Concept-based 3D Constitutive Models for Geomaterials*, Taylor & Francis, New York.

McDowell, G. R., and Hau, K. W. (2003). "A Simple Non-associated three surface kinematic hardening model." *Geotechnique*, 53(4), 433-437.

Mesri, G., and Castro, A. (1987). "$C_\alpha/C_c$ Concept and $K_0$ During Secondary Compression." *Journal of Geotechnical Engineering*, 113(3), 230-247.

Mroz, Z. (1967). "On the description of anisotropic work hardening." *Journal of Mechanics and Physics of Solids,* 15, 163-175.

Murakami, Y. (1979). "Excess pore-water pressure and preconsolidation effect developed in normally consolidated clays of some age." *Soils and Foundations*, 19(4), 17-29.

Nash, D. F. T. (2001). "Discussion: Precompression design for secondary settlement reduction." *Geotechnique*, 51(9), 822–823.

Oka, F., Adachi, T., and Okano, Y. (1986). "Two-dimensional consolidation analysis using an elasto-viscoplastic constitutive equation." *International Journal for Numerical and Analytical Methods in Geomechanics*, 10(1), 1-16.

Perzyna, P. (1963). "Constitutive equations for rate-sensitive plastic materials." *Quarterly of Applied Mathematics*, 20, 321-331.

Prashant, A., and Penumadu, D. (2005). "A laboratory study of normally consolidated kaolin clay." *Canadian Geotechnical Journal*, 42(1), 27-37.

Roscoe, K. H., and Schofield, A. N. (1963)."Mechanical behavior of an idealised 'wetclay' " , *Proc. of European Conference of Soil Mechanics and Foundation Engineering*, 47-54.

Roscoe, K. H., and Burland, J. B. (1968). "On the generalized stress-strain behavior of wet clay." *In Engineering plasticity*, , J. Heyman, and F. A. Leckie, eds., Cambridge University Press, Cambridge, UK, , 535-609.

Sheng, D., Sloan, W. S., and Yu, S. H. (2000). "Aspects of finite element implementation of critical state models." *Computational Mechanics*, 26(2), 185-196.

Willam, K. J., and Warnke, E. P. (1975). "Constitutive model for the triaxial behaviour of concrete." *ISMES Seminar on Concrete Structures Subjected to Triaxial Stress*, Bergamo, Italy, 1-30.

Whittle, A., and Kavvadas, M. (1994). "Formulation of MIT-E3 Constitutive Model for Overconsolidated Clays." *Journal of Geotechnical Engineering*, 120(1), 173-198.

Xiao, Y., Sun, Y., Liu, H., and Yin, F. (2016). "Critical state behaviors of a coarse granular soil under generalized stress conditions." *Granular Matter*, 18(2), 17.

Yao, Y.-P., and Sun, D. A. (2000). "Application of Lade's Criterion to Cam-Clay Model." *Journal of Engineering Mechanics*, 126(1), 112-119.

Yao, Y.-P., and Wang, N.-D. (2014). "Transformed Stress Method for Generalizing Soil Constitutive Models." *Journal of Engineering Mechanics*, 140(3), 614-629.

Yao, Y.-P., Kong, L.-M., Zhou, A.-N., and Yin, J.-H. (2015). "Time-Dependent Unified Hardening Model: Three-Dimensional Elastoviscoplastic Constitutive Model for Clays." *Journal of Engineering Mechanics*, 141(6), 04014162.

Ye, G.-L., Ye, B., and Zhang, F. (2014). "Strength and Dilatancy of Overconsolidated Clays in Drained True Triaxial Tests." *Journal of Geotechnical and Geoenvironmental Engineering*, 140(4), 06013006.

Yin, J. H. (1999). "Non-linear creep of soils in odeometer tests." *Geotechnique*, 49(5), 699-707.

Yin, J. H., and Zhu, J. G. (1999). "Measured and predicted time-dependent stress-strain behaviour of Hong Kong marine deposits." *Canadian Geotechnical Journal*, 36(4), 760-766.

Yin, J. H., Zhu, J. G., and Graham, J. (2002). "A new elastic viscoplastic model for time-dependent behaviour of normally and overconsolidated clays: theory and verification." *Canadian Geotechnical Journal*, 39(1), 157-173.

Yin, Z., Xu, Q., and Yu, C. (2015). "Elastic-Viscoplastic Modeling for Natural Soft Clays Considering Nonlinear Creep." *International Journal of Geomechanics*, 15(5), A6014001-6014010.

Yu, H. S. (2006). "*Plasticity and Geotechnics*", Springer, USA.

Zienkiewicz, O. C., Humpheson, C., and Lewis, R. W. (1975). "Associated and non-associated visco-plasticity and plasticity in soil mechanics." *Geotechnique*, 25(4), 671-689.

Zhu, J. G. (2000). "Experimental study and elastic visco-plastic modelling of the time-dependent stress-strain behaviour of Hong Kong marine deposits." Dept. of Civil and Structural Engineering, The Hong Kong Polytechnic University, PhD Thesis, Hong Kong.

**List of Figures:**

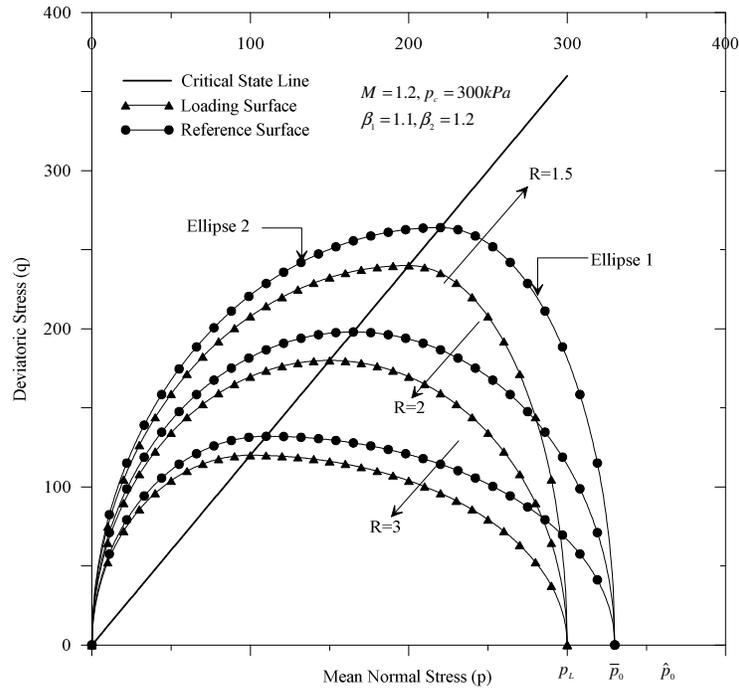

Figure 1: Illustration of reference and loading surfaces in *p-q* plane

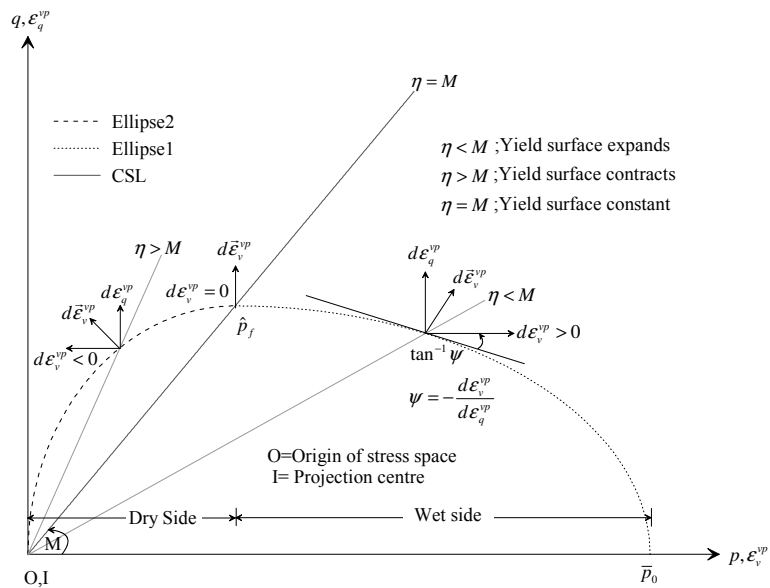

Figure 2: Meridional section of reference surface for two ellipses

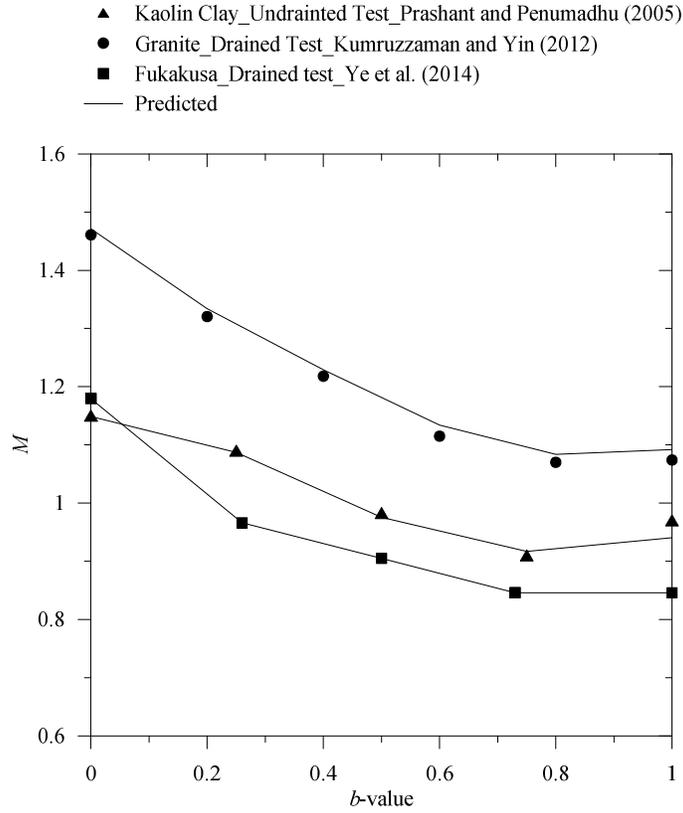

Figure 3: Relation between *M* and *b*-value

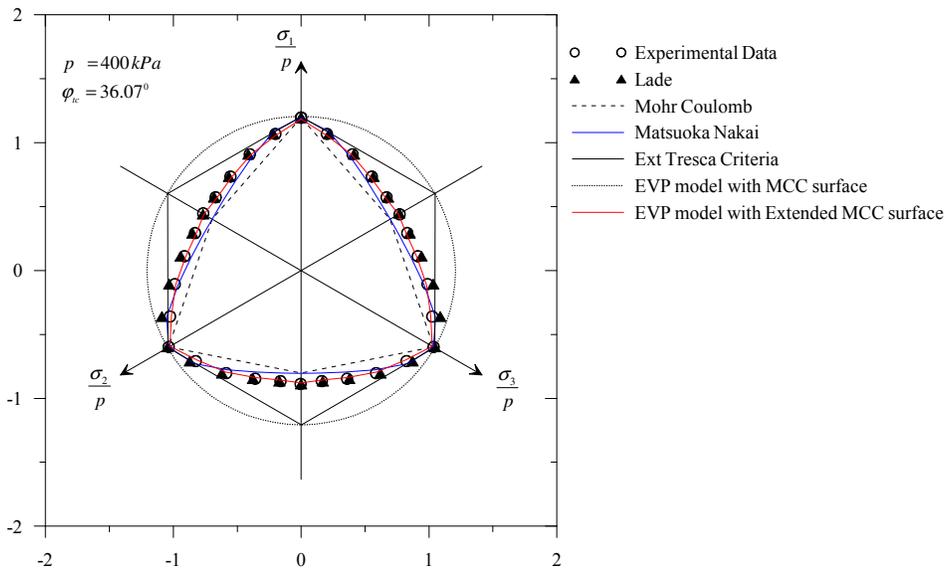

Figure 4: Three dimensional failure surface in an octahedral plane

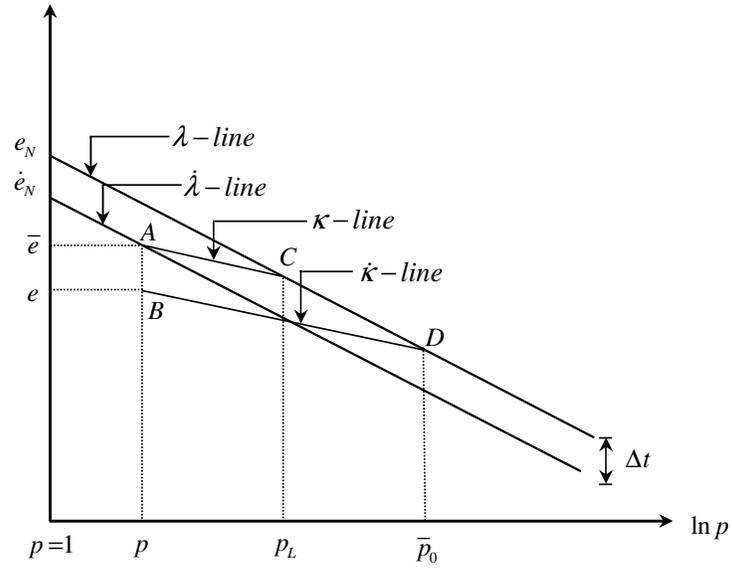

Figure 5: Relative locations of $p_L$ and $\bar{p}_0$ in *e-lnp* space

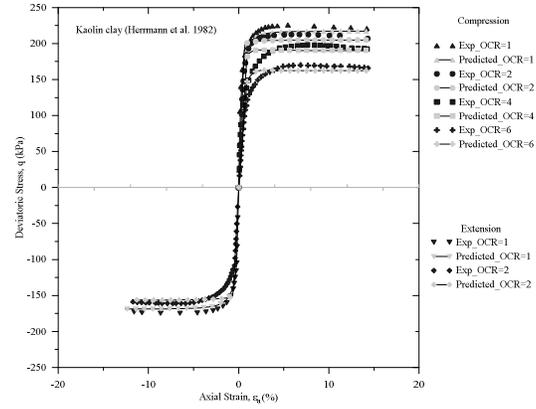

(a)

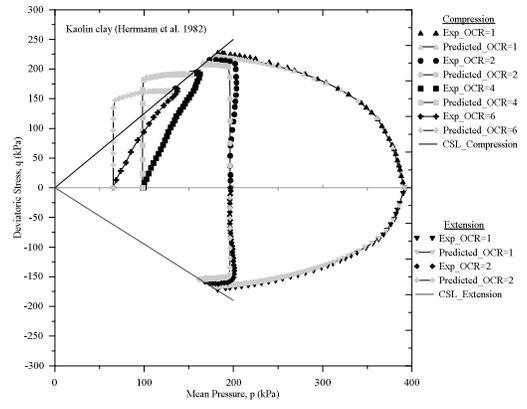

(b)

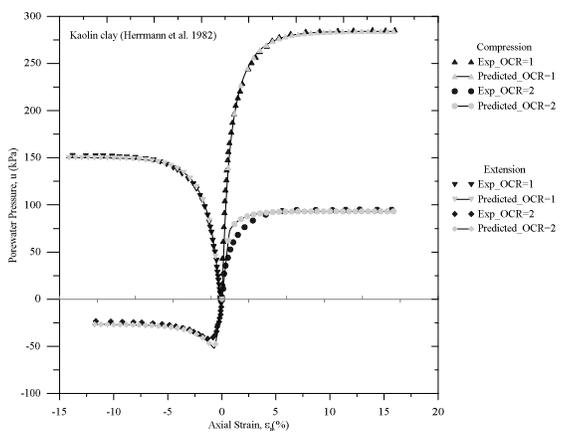

(c)

Figure 6: Comparison of measured and predicted consolidated undrained triaxial compression tests results on Kaolin clay (data from Herrmann et al. 1982): (*a*) deviatoric stress-axial strain, (*b*) stress path, and (*c*) pore-water pressure-axial strain

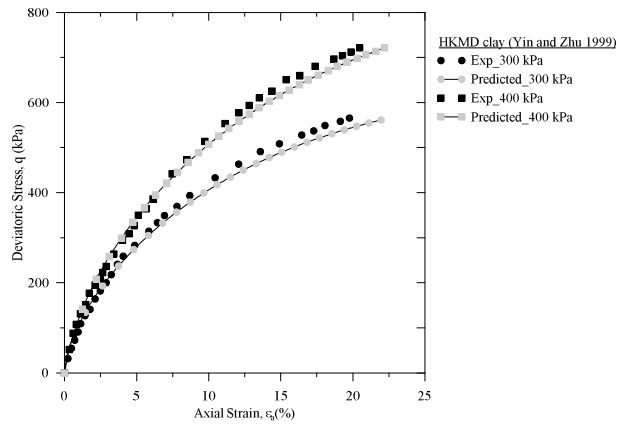

(a)

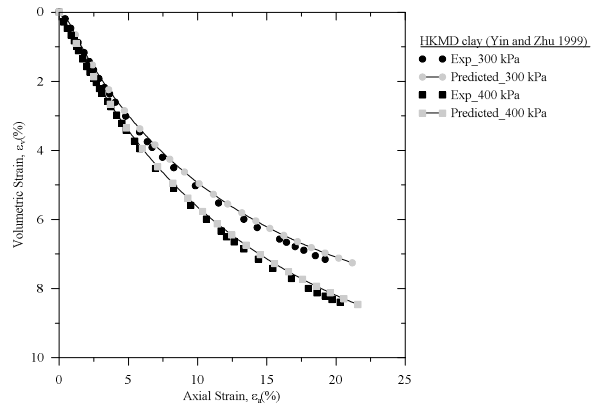

(b)

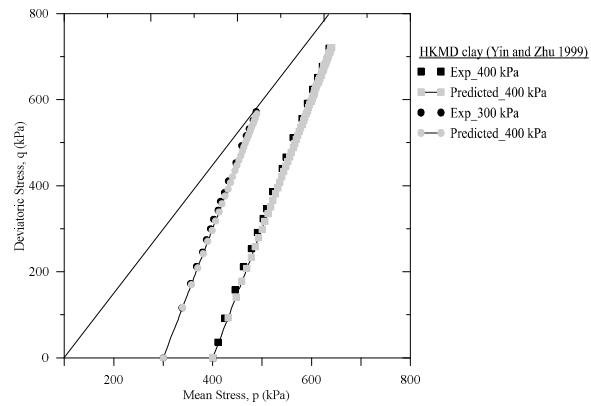

(c)

Figure 7: Comparison of measured and predicted consolidated drained triaxial compression tests results on HKMD clay (data from Yin and Zhu 1999): (*a*) deviatoric stress-axial strain, (*b*) volumetric strain-axial strain, and (*c*) stress path

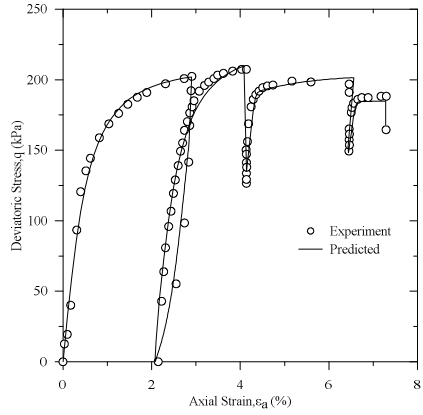

(a)

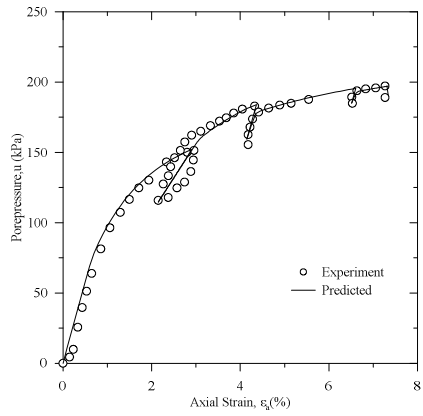

(b)

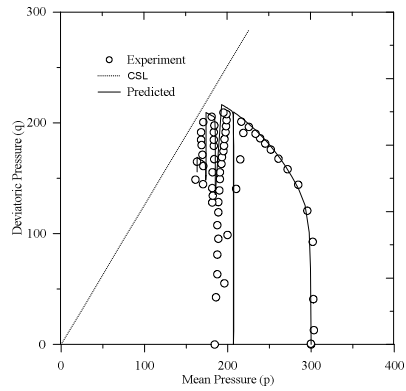

(c)

Figure 8: Comparison of measured and predicted undrained triaxial tests results for stage-changed axial strain rate combined with stress relaxation on HKMD clay (data from Yin et al. 2002): (*a*) deviatoric stress-axial strain, (*b*) pore pressure-axial strain stress path, and (*c*) stress path

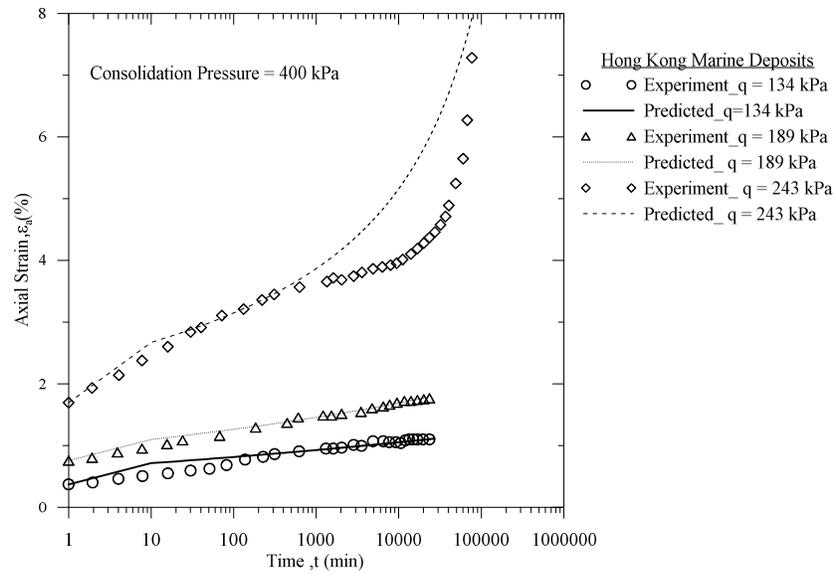

(a)

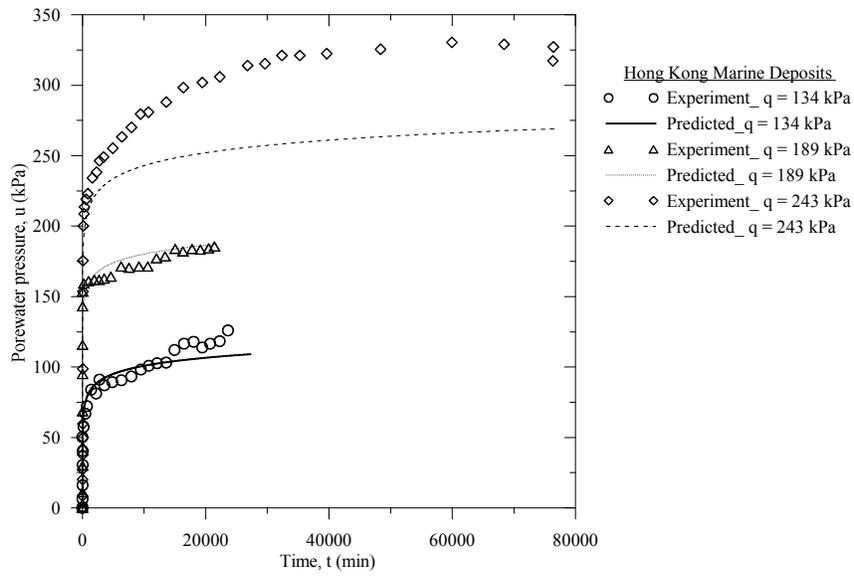

(b)

Figure 9: Comparison of measured and predicted undrained triaxial creep tests results on HKMD clay (data from Zhu 2000): (*a*) axial strain and time relation, and (*b*) pore pressure and time relation

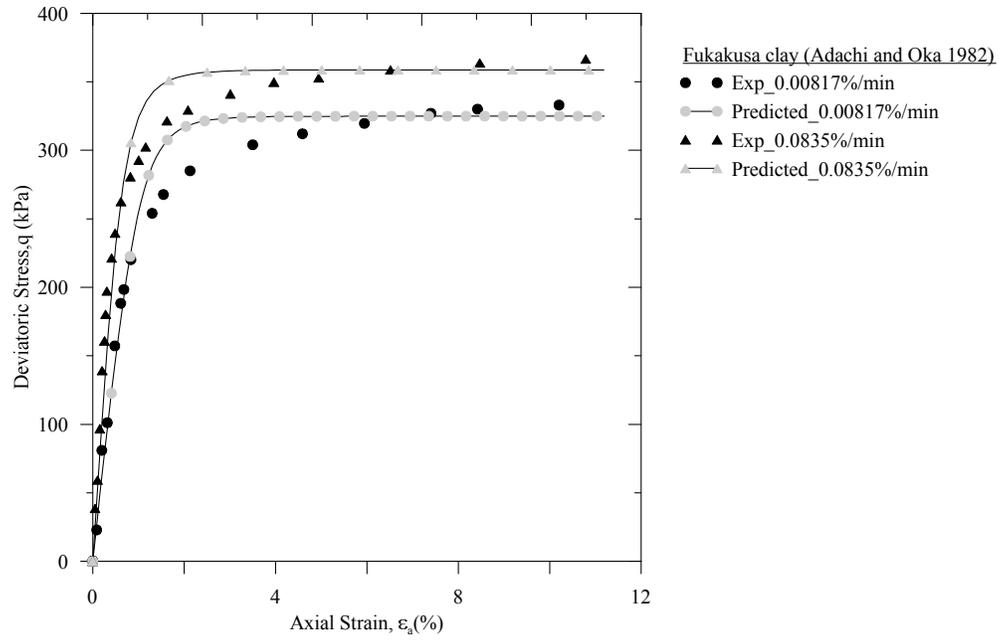

(a)

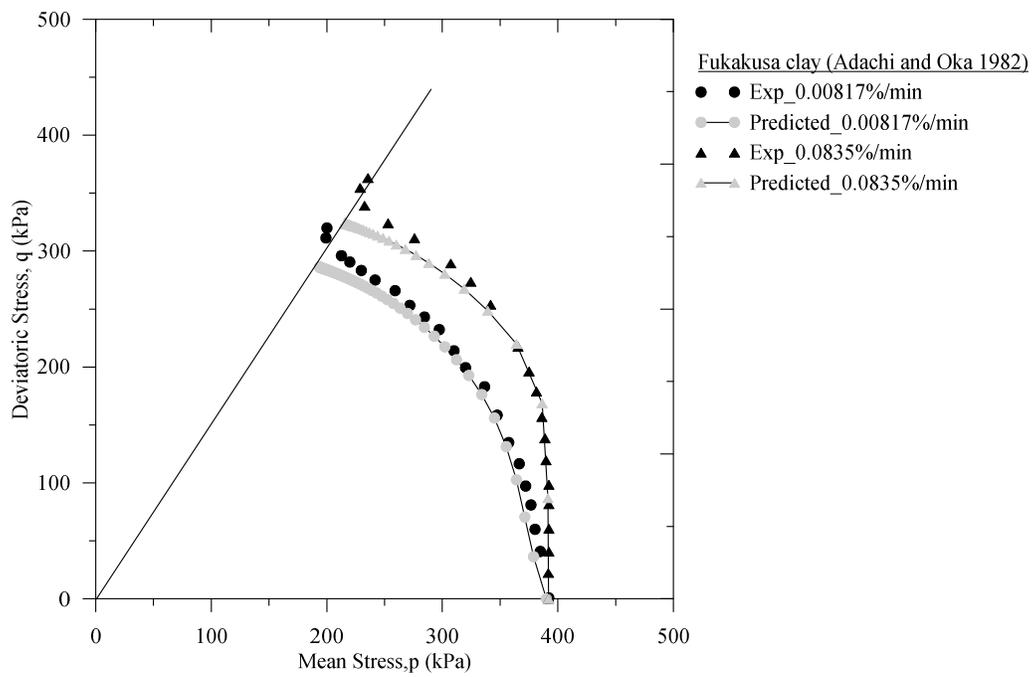

(b)

Figure 10: Comparison of measured and predicted undrained triaxial tests results on Fukakusa clay (data from Adachi and Oka 1982): (*a*) deviatoric stress- axial strain, and (*b*) stress path

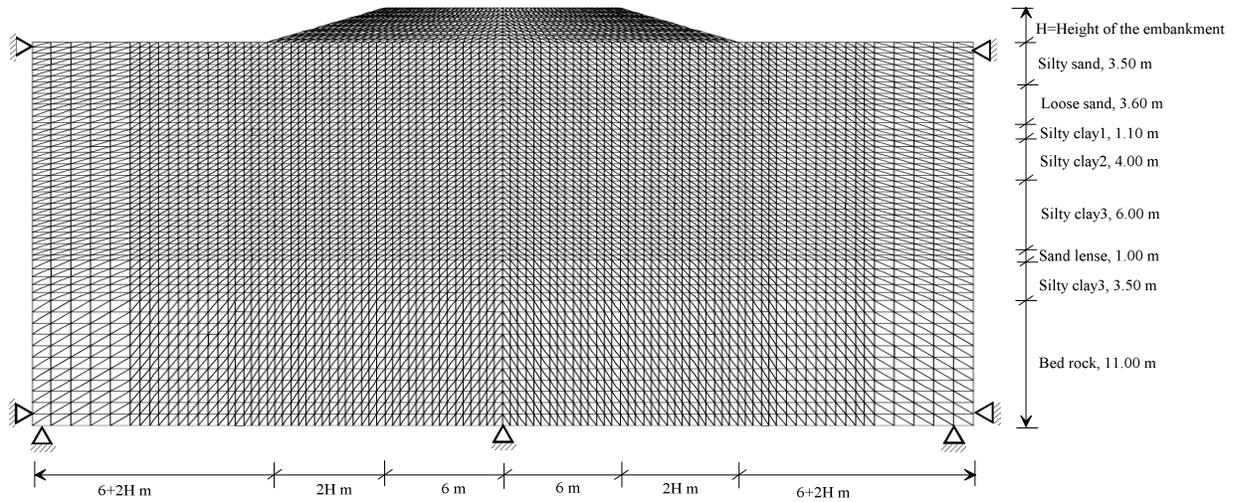

Figure 11: Finite element geometry for plane strain analysis

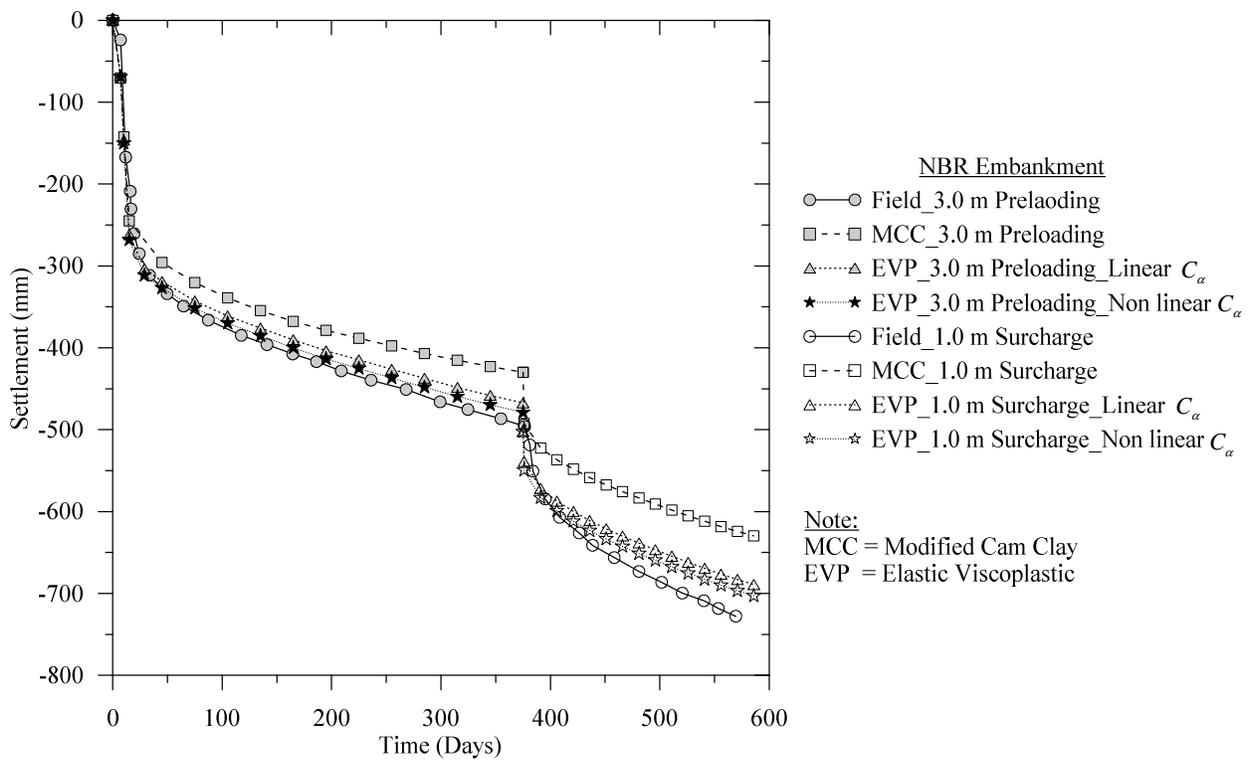

Figure 12: Predicted and measured settlement for NBR embankment

Table 1: Parameters used for prediction of laboratory data

| Soil Types | Model Properties | | | | | | | | $K_i$ cm/min | Test Types | References |
|---|---|---|---|---|---|---|---|---|---|---|---|
| | $\lambda$ | $\kappa$ | $M_c$ | $M_e$ | $\nu$ | $e_N$ | $C_\alpha$ | R | | | |
| Kaolin Clay | 0.15 | 0.018 | 1.25 | 0.95 | 0.3 | 1.515 | 0.0139 | 2.50 | $3.2 \times 10^{-7}$ | CUC &CUE) | Hermann et al. (1982) |
| HKMD Clay | 0.20 | 0.045 | 1.26 | 0.89 | 0.3 | 2.187 | 0.0106 | 2.00 | $3.7 \times 10^{-6}$ | CUC,CDC, CU(L-U-RL) | Zhu (2000), Yin and Zhu (1999), Yin et al. (2002) |
| Fukakusa Clay | 0.10 | 0.02 | 1.50 | --- | 0.3 | 1.31 | 0.0064 | 2.38 | $1.6 \times 10^{-7}$ | CUC | Adachi and Oka (1982) |

*Note:* HKMD - Hong Kong Marine Deposit, CUC - Consolidated Undrained Compression, CUE - Consolidated Undrained Extension,

CDC - Consolidated Drained Compression, L-U-RL - Loading-Unloading-Reloading, $M_c$ and $M_e$ for triaxial compression and extension respectively.

Table 2: Material parameters used in the prediction of NBR embankment field data

| RL (m) | Soil Layer | M | λ | κ | $e_N$ | $p_c^*$ kPa | $C_\alpha$ | $k_v$ m/day | $K_0$ |
|---|---|---|---|---|---|---|---|---|---|
| +1.19 to 1.19+H | Fill | \multicolumn{6}{c}{$E' = 3000$ kPa, $\varphi' = 30°$, $c' = 5.0$ kPa, $k = 250$, $n = 0.5$} | | --- | 0.50 |
| +1.19 to -2.31 | Silty sand | \multicolumn{6}{c}{$E' = 5000$ kPa, $\varphi' = 35°$, $c' = 2.5$ kPa, $k = 250$, $n = 0.5$} | | --- | 0.43 |
| -2.31 to -5.90 | Loose sand | \multicolumn{6}{c}{$E' = 7000$ kPa, $\varphi' = 33°$, $c' = 1.5$ kPa, $k = 250$, $n = 0.5$} | | --- | 0.46 |
| -5.90 to -7.00 | Silty clay-1 | 1.28 | 0.36 | 0.060 | 2.10 | 159.52 | 0.029 | 2.65× $10^{-5}$ | 0.47 |
| -7.00 to -11.00 | Silty clay-2 | 1.25 | 0.42 | 0.043 | 3.73 | 105.36 | 0.033 | 2.16× $10^{-5}$ | 1.48 |
| -11.00 to -17.00 | Silty clay-3 | 1.20 | 0.29 | 0.030 | 2.61 | 132.20 | 0.023 | 1.03× $10^{-5}$ | 0.50 |
| -17.00 to -18.00 | Sand lense | \multicolumn{6}{c}{$E' = 3000$ kPa, $\varphi' = 35°$, $c' = 5.0$ kPa, $k = 250$, $n = 0.5$} | | --- | 0.43 |
| -18.00 to -21.50 | Silty clay-3 | 1.20 | 0.29 | 0.030 | 2.61 | 287.18 | 0.023 | 1.03× $10^{-5}$ | 0.50 |
| -21.50 to -32.50 | Bed rock | \multicolumn{6}{c}{$E' = 15000$ kPa, $\varphi' = 36°$, $c' = 50.0$ kPa, $k = 250$, $n = 0.5$} | | --- | 0.41 |

*Notes:* Poisson's ratio considered 0.3

* At top of soil layer

$K_0$ = lateral earth pressure. For normally consolidated clay, $K_0 = 1 - sin\varphi$, whereas for over consolidated clay $K_0 = (1 - sin\varphi) OCR^{-\frac{1}{2}}$. H = filling height; RL = reduced level.